\documentclass[11pt,a4paper]{article}
\pdfoutput=1
\usepackage{amsmath,amssymb,amsfonts,a4wide,graphicx,bm,times,psfrag,wrapfig,sidecap}
\usepackage{cite}
\usepackage[colorlinks=true,linkcolor=black, citecolor=black,
urlcolor=black]{hyperref} 
\numberwithin{equation}{section}
\makeatletter \let\old@startsection=\@startsection
\renewcommand{\@startsection}[6]
{\old@startsection{#1}{#2}{#3}{#4}{#5}{#6\mathversion{bold}}}
\makeatother

\def\defeq{\stackrel{\text{def}}=}
\newcommand\re[1]{({\ref{#1}})}
\def\be{\begin{eqnarray}}
    \def\ee{\end{eqnarray}}
 \def\z{\zeta}
 \def\IR{{\mathbb{R}}}
\def\IZ{{\mathbb{Z}}}
 
 \def\IN{{\mathbb{N}}}

    \def\no{\nonumber}
    \def\la{\label}
    \def\l {\lambda}
\def\({\left(} \def\){\right)} 
\def\<{\left\langle\,} 
\def\>{\, \right\rangle} 
\def\[{\left[}
 \def\]{\right]} 
\def\tr{{\rm   tr} }
    \def\hf{ {\textstyle{1\over 2}} }
    \def\CA{{\cal A}}
 
\def\CO{{ \mathcal{ O} }}
    \def\CZ{{\cal Z}}
       \def\CN{{  N}}
       
  \def\CH{ {\cal H}}
  \def\CU{{\cal U}}
   \def\CG{{\cal G}}
 \def\CF{{\cal F}}
  \def\p{\partial}
  \def\a{\alpha}
 \def\b{\beta}
  \def\g{\gamma}
  \def\e{\epsilon}
  \def\s{\sigma}
  \def\t{\tau}
  \def\th{\theta}
  \def\G{\Gamma}
 \def\reg{{\rm reg}}
  
 \def\Tr{{\rm Tr}}
 \def\mb{ \mu_{_{\text{B}}}} 
 \def\bmb{ {\bar \mu}_{_{\text{B}}}} 
 \def\YL{ {_{^ {\text{YL}}}}}
\def\cor{\text{cor}}
\def\sp{ {\text{sphere}}}
\def\di{{\text{disk}}}
 \def\bmu{ \bar\mu}
 \def\Fb{ { {\CF}_{\text{B}}}}

\def\high{{\text{high}}}
\def\low{{\text{low}}}
\def\CGh{{\CG}_\high}
\def\CGl{{\CG}_\low}
\def\OnFT{ {O_n \text{FT}}}
\def\OnQG{ {O_n \text{QG}}}
\def\d{\delta}
\def\kappa{\omega }
 \def\s{{\rm s}}

\begin{document}

\thispagestyle{empty}

\begin{flushright}
  IPhT/t11/195  
\end{flushright}

\vspace{1cm}
\setcounter{footnote}{0}

\begin{center}

   {\Large\bf On the Yang-Lee and Langer singularities in the $O(n)$
   loop model }

\vspace{20mm} 

Jean-Emile Bourgine$^\star$ and Ivan Kostov$^\ast$\footnote{\it
Associate member of the Institute for Nuclear Research and Nuclear
Energy, Bulgarian Academy of Sciences, 72 Tsarigradsko Chauss\'ee,
1784 Sofia, Bulgaria} \\[7mm]

{\it $^\star$ Center for Quantum Spacetime (CQUeST)\\ 
Sogang University, Seoul 121-742, Korea} \\[5mm]

{\it $^\ast$ Institut de Physique Th\'eorique, CNRS-URA 2306 \\
	     C.E.A.-Saclay, \\
	     F-91191 Gif-sur-Yvette, France} \\[5mm]

\end{center}

\vskip9mm

\vskip18mm

\noindent{ We use the method of `coupling to 2d QG' to study the
analytic properties of the universal specific free energy of the
$O(n)$ loop model in complex magnetic field.  We compute the specific
free energy on a dynamical lattice using the correspondence with a
matrix model.  The free energy has a pair of Yang-Lee edges on the
high-temperature sheet and a Langer type branch cut on the
low-temperature sheet.  Our result confirms a conjecture by A. and Al.
Zamolodchikov about the decay rate of the metastable vacuum in
presence of Liouville gravity and gives strong evidence about the
existence of a weakly metastable state and a Langer branch cut in the
$O(n)$ loop model on a flat lattice.  Our results are compatible with
the Fonseca-Zamolodchikov conjecture that the Yang-Lee edge appears as
the nearest singularity under the Langer cut.  }

\newpage
\setcounter{page}{1}
\setcounter{footnote}{0}
%
\section{Introduction and summary}
\label{sec:Introduction}

In the theory of phase transitions, the relevance of the analytic
continuation of the free energy to complex values of its parameters is
known since the work of Lee and Yang \cite{Yang:1952be}.  They have
shown that the thermodynamical equation of state is completely
determined by the distribution of zeros of the partition function in
the fugacity complex plane, or in the magnetic field complex plane in
the case of spin systems.  The distribution of zeros determine the
analytical properties of the free energy -- they typically accumulate
in arrays and produce branch cuts.  From the analytical properties of
the free energy one can reconstruct the qualitative picture of the
critical phases and the phase boundaries.

The most studied example is the ferromagnetic Ising model in a complex
magnetic field.  In the high-temperature phase ($T>T_\text{crit}$),
the free energy as a function of the magnetic field has two branch
cuts, starting at the points $H_\YL^\pm = \pm i H_c(T)$ on the
imaginary axis\cite{PhysRev.87.410}.  The gap between the two branch
points vanishes at the critical temperature $T_\text{crit}$.  Fisher
\cite{PhysRevLett.40.1610} named the branch points Lee-Yang
singularities, and suggested that they are described by a non-unitary
field theory with a cubic potential.  Cardy \cite{Cardy:1985yy}
identified the Lee-Yang singularity as the simplest non-unitary CFT,
the minimal model $\mathcal{M}_{2,5}$ with a central charge $c_\YL = -
22/5$.  On the other hand, in the low-temperature phase
($T<T_\text{crit}$), the free energy has a weak singularity at the
origin, predicted by Langer's theory \cite{langer1967theory,
Langer:1969aa}.  The analytical continuation from the positive part of
the real axis has a branch cut on the negative axis, known as Langer
branch cut.  The discontinuity across this cut is exponentially small
when $T\to T_\text{crit}$.  It can be explained with condensation of
droplets of the stable phase near a metastable point.

A thorough analysis of the analytic properties of the free energy near
the critical point $(H=0, T=T_\text{crit})$, based on the truncated
CFT approach \cite{Yurov:1991my}, was performed by Fonseca and
Zamolodchikov \cite{Fonseca:2001dc}.  In the continuum limit, the
Ising model is described by an euclidean quantum field theory, the
Ising field theory (IFT), which is a double perturbation of massless
Majorana fermion.  The Ising field theory is formally defined by the
action
\be\la{ActIsingFT} \CA_{\rm IFT} = \CA_{c=1/2} - t \int \varepsilon(x)
\, d^2 x - h \int \s (x) \, d^2 x, \ee
where the first term is the action of a free massless Majorana
fermion, $\s(x)$ is the spin field with conformal weights $\Delta_\s=
\bar\Delta_\s=1/16$ and $\varepsilon$ is the energy density with
conformal weight $\Delta_\varepsilon = \bar\Delta_\varepsilon =1/2$.
The coupling constants $t$ and $h$ are the renormalized temperature
and magnetic field, they parametrize the vicinity of the critical
point $T=T_\text{crit}, H=0$.  In the QFT language, the specific free
energy is interpreted as the vacuum energy density, and its
discontinuity along the Langer branch cut as the decay rate of a
``false vacuum'' \cite{PhysRevD.15.2929,Callan:1977pt}.  By
dimensional arguments, the vacuum energy density of the Ising field
theory must have the form
\be\la{defGxi}
F(t, h) \sim t^2 \log t ^2 + t^2 \, G(\xi), 
\ee
where the first term is the free energy of a (massive) Majorana
fermion and $G(\xi)$ is a scaling function of the dimensionless
strength of the magnetic field,
\be\la{defxiflat} \xi = h/|t|^{15/8}.  \ee
The high temperature ($t>0$) and the low-temperature ($t<0$) regimes
are described by two different scaling functions,
$G_{\text{high}}(\xi)$ and $G_{\text{low}}(\xi)$.

\begin{figure}
         \centering
         \begin{minipage}[t]{0.4\linewidth}
            \centering
            \includegraphics[width=4.5cm]{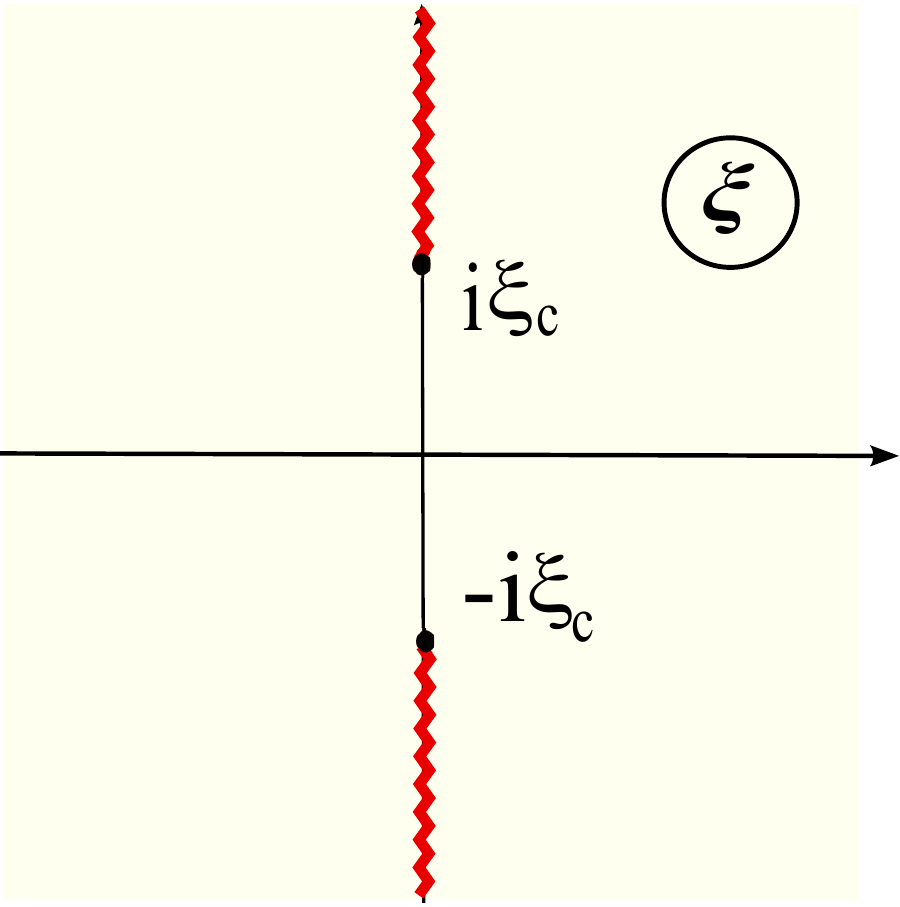}
 \caption{ \small Analytic structure of $G_\high (\xi)$ in the complex
 $\xi$-plane.  The function has two singularities (YL edges) at $\xi=
 \pm i \xi_c$ and a third one at $\xi=\infty$ (the critical point).
 There are two cuts along the imaginary axis connecting the two YL
 edges and the critical point.  }
  \label{fig:HTcuts}
         \end{minipage}%
         \hspace{2cm}%
         \begin{minipage}[t]{0.4\linewidth}
            \centering
            \includegraphics[width=4.5cm]{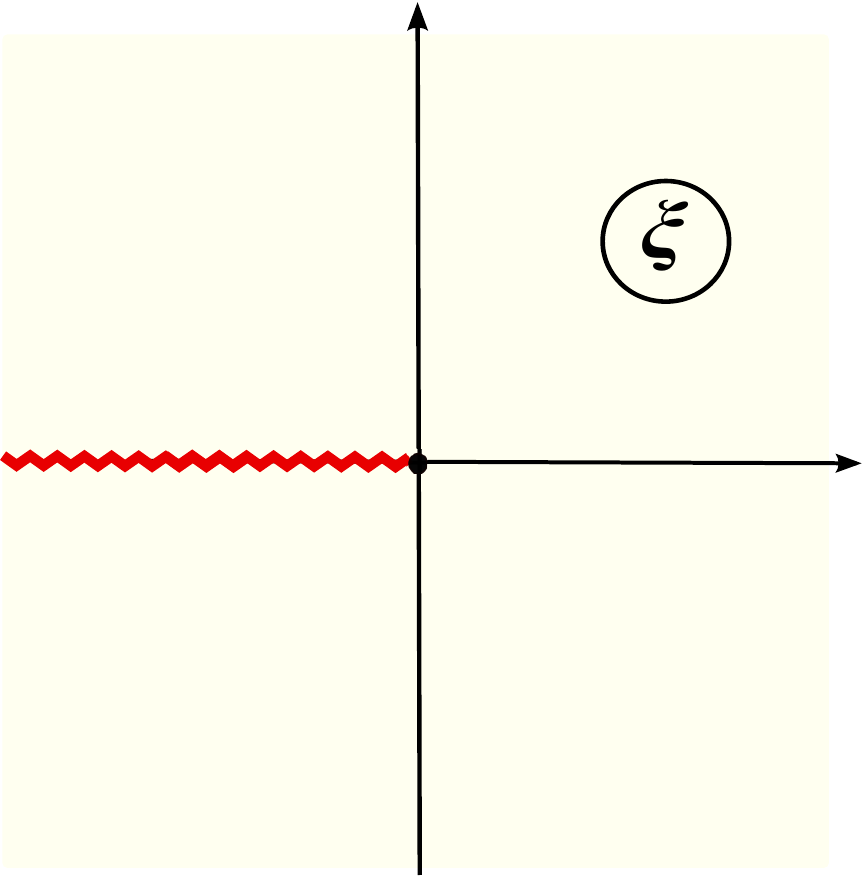}
	     \caption{\small Analytic structure of $G_\low(\xi)$.  The
	     function has two singularities, the critical point at
	     $\xi=\infty$ and the low-temperature fixed point at
	     $\xi\to 0$.  The two singularities are connected by a
	     branch cut extending from $-\infty$ to 0.}
\label{fig:LTcuts}
         \end{minipage}
 \vskip 1cm
      \end{figure}

The function $G_{\high}(\xi)$ is analytic at small $\xi$ and has two
symmetric Yang-Lee branch points at $\xi = \pm i \xi_c$ on the
imaginary axis (Fig.  \ref{fig:HTcuts}), while the function
$G_{\low}(\xi)$ is characterized by Langer branch cut starting at the
origin (Fig.  \ref{fig:LTcuts}).  The two scaling functions are
analytically related in the vicinity of the point $\xi=\infty$ because
the magnetic field creates a mass gap and smears the phase transition
at $t=0$.  Therefore $G_{\high}$ and $G_{\low}$ must correspond to two
different branches of the same multivalued function of the variable
$\xi$.  These analytic properties represent the standard analyticity
assumptions; they do not exclude the presence of more singularities in
the other sheets of $G(\xi)$.

The numerical results obtained in \cite{Fonseca:2001dc} led the
authors to a stronger assumption, which they referred to as `extended
analyticity', and which states that the Yang-Lee edge is the {\it
nearest singularity} under the Langer branch cut.  The extended
analyticity suggests an interpretation of the YL edge as a quantum
counterpart of the { spinodal point} in the classical theory of
phase transitions.  The classical low-temperature free energy is
regular at $H=0$, but shows a branch-cut singularity at some negative
$H= - H_{\rm SP}$, where the metastable phase becomes classically
unstable.  It was argued in \cite{Fonseca:2001dc} that the spinodal
singularity does not disappear completely due to the quantum effects,
but moves under the Langer cut and reappears as YL edge in the
high-temperature phase.  More recently, the numerical results of
\cite{Fonseca:2001dc} were confirmed by a calculation based on the
lattice formulation of the problem \cite{Mangazeev:2009rm}.

 It is tempting to think that the qualitative picture of the analytic
 properties of the free energy, proposed in \cite{Fonseca:2001dc},
 applies to all statistical systems having a line of first order
 transitions ending at a second order transition point, and that the
 extended analyticity conjecture has a universal character.  However,
 it is not clear how to address this issue directly.  The Langer
 singularity is too weak to to be measured experimentally, and the
 TCFT approach used in \cite{Fonseca:2001dc} is sufficiently reliable
 only in the case of Majorana fermions.\footnote{An attempt to use the
 approach of of \cite{Fonseca:2001dc} to study the tricritical Ising
 model in magnetic field was made in \cite{Mossa:2008aa}.  }

On the other hand, it is known that some questions about the
two-dimensional statistical systems can be answered by the following
detour: formulate the system on a dynamical lattice, solve the problem
exactly and then interpret the answer for the original system.  This
procedure is usually called `coupling to 2D quantum gravity'.  Due to
the enhanced symmetry, which erases the coordinate dependence of the
correlation functions, the systems on dynamical lattices are much
simple to resolve.  This approach is based on the fact that there is a
well established correspondence between the universal properties on
the flat and dynamical lattices.  On the dynamical lattice, the
description of the critical points is given in terms of Liouville
gravity \cite{Polyakov:1981rd}.  The matter in Liouville gravity is
given by the CFT describing the critical point of the statistical
system on a flat lattice, while the Liouville field describes the
fluctuations of the metric.  The matter fields influence the geometry
in a precise way, which is taken into account by accompanying
Liouville vertex operators \cite{Knizhnik:1988ak, David:1988hj,
distler1989conformal}.  Knowing the correlation functions in 2D
gravity, one can reconstruct the scaling dimensions of the matter
operators.  Moreover, the method based on coupling to 2D gravity is
efficient also away from the critical points, where it makes possible
to reconstruct the bulk and the boundary renormalization flow diagrams
of the matter theory on a flat lattice, knowing the exact solution on
dynamical lattice.  The corresponding field theory is Liouville
gravity perturbed by one or several fields.

The ising model is the only nontrivial example where the analytic
properties of the free energy are known both on flat and dynamical
lattices.  The exact solution of Boulatov and Kazakov
\cite{Boulatov:1986sb} shows that the Yang-Lee singularities appear
also on a dynamical lattice and that near each YL singularity the
critical behavior is that of a Yang-Lee CFT coupled to gravity
\cite{Staudacher:1989fy,Crnkovic:1989tn,Brezin:1989db}.  However, it
is not {a priori} clear if coupling to 2D gravity could be used to
study the decay of the metastable vacuum by nucleation.  Such a
possibility was first explored in an unpublished work by Al.
Zamolodchikov \cite{AlZ-unp} and a subsequent work by A. and Al.
Zamolodchikov \cite{Zamolodchikov:2006xs}.

The universal behavior of the Ising model on dynamical lattice is
described by Ising Quantum Gravity (IQG), which represents a double
perturbation of Liouville gravity with matter central charge $1/2$.
The IQG is formally defined by the action
    \begin{eqnarray}\label{actgIsi}
{\cal A}_{\text{IQG} }= \CA_{ c=1/2\ \text{Liouville gravity}} - t \,
\int {\bf \varepsilon}(x) \, e^{2\a_\varepsilon \phi (x) } d^2 x - h
\, \int {\bf \sigma}(x) \, e^{2 \a_\sigma \phi(x)} d^2 x.
\end{eqnarray}
The exponents of Liouville dressing operators, $\a_\varepsilon = 1/3$
and $\a_\sigma = 5/6$, are such that the dressed matter operators
become densities \cite{David:1988hj,distler1989conformal}.  The vacuum
energy density of IQG must be of the form
  \be
  \la{defscalingfd}
   \CF_{\text{IQG}}(t, h)= t^3 \CG(\xi), \qquad \xi = h/ t ^{5/2}, \ee
where the exponent $5/2= \a_\sigma/\a_\varepsilon $ is such that $\xi$
is invariant under rescalings of the metric.  The vacuum energy
density of the IQG, or the universal specific free energy of the
discrete model, was extracted in \cite{AlZ-unp} from the
Bulatov-Kazakov exact solution \cite{Boulatov:1986sb}.  For some
normalization of the coupling constants, the universal free energy is
given in parametric form by
       \be
       \la{FreeIQG}
       \begin{aligned}
       \CF_{\text{IQG}}(t, h)& = - \hf \, u^3 - \textstyle{3\over 4}
       \, u^2 - \textstyle{3\over 4} \, h^2\, (u+t)^{-2} \, , \\
    h^2\ \   &=  u\, (u+t)^4 .
     \end{aligned}
       \ee

The scaling function $ \CG(\xi)$ defined in \re{defscalingfd} is
analytic on a Riemann surface representing a 5-sheet covering of the
complex $\xi$-plane.  The Riemann surface has a high-temperature (HT)
sheet with two YL branch points at $\xi = \pm i \xi_c$ and two copies
of the low-temperature (LT) sheet having a single cut starting at
$\xi=0$ and going to infinity.  The HT sheet is connected to the two
LT sheets by two auxiliary sheets.  In this way, the scaling functions
$\CG_\high$ and $\CG_\low$ are identified with the restrictions of a
single function $\CG$ to the HT and the LT sheet of its Riemann
surface.

\begin{figure}
         \centering
         \begin{minipage}[t]{0.4\linewidth}
            \centering
            \includegraphics[width=4.4cm]{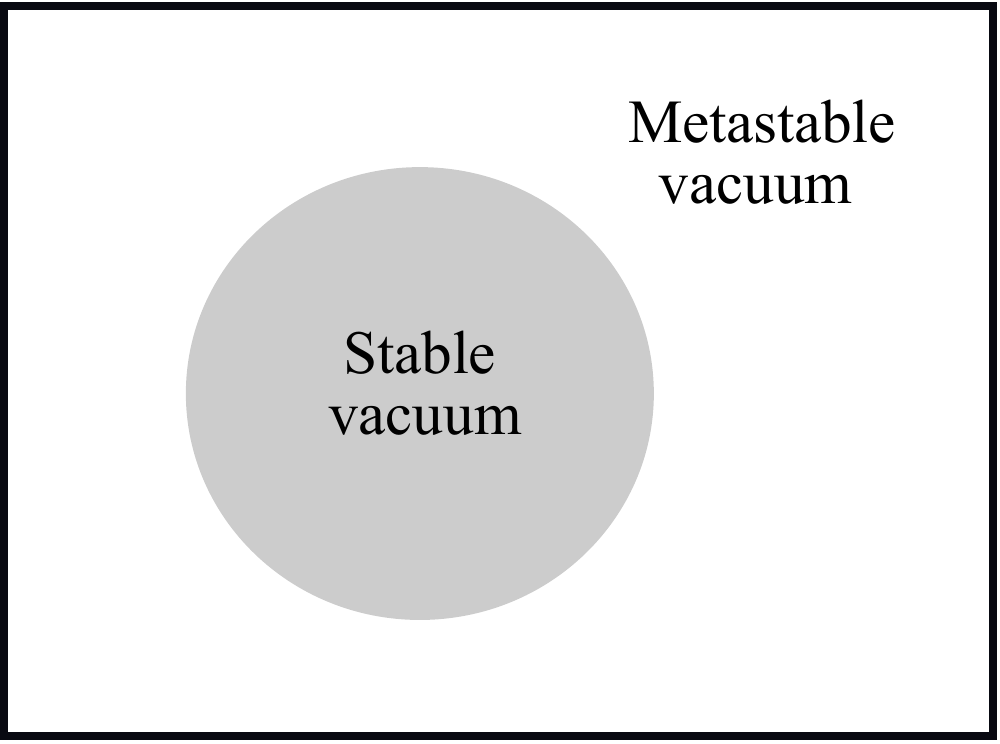}
 \caption{ \small A critical droplet of the true vacuum in the sea of
false vacuum. } \label{fig:Droplet}
         \end{minipage}%
         \hspace{2cm}%
         \begin{minipage}[t]{0.4\linewidth}
            \centering
            \includegraphics[width=6cm]{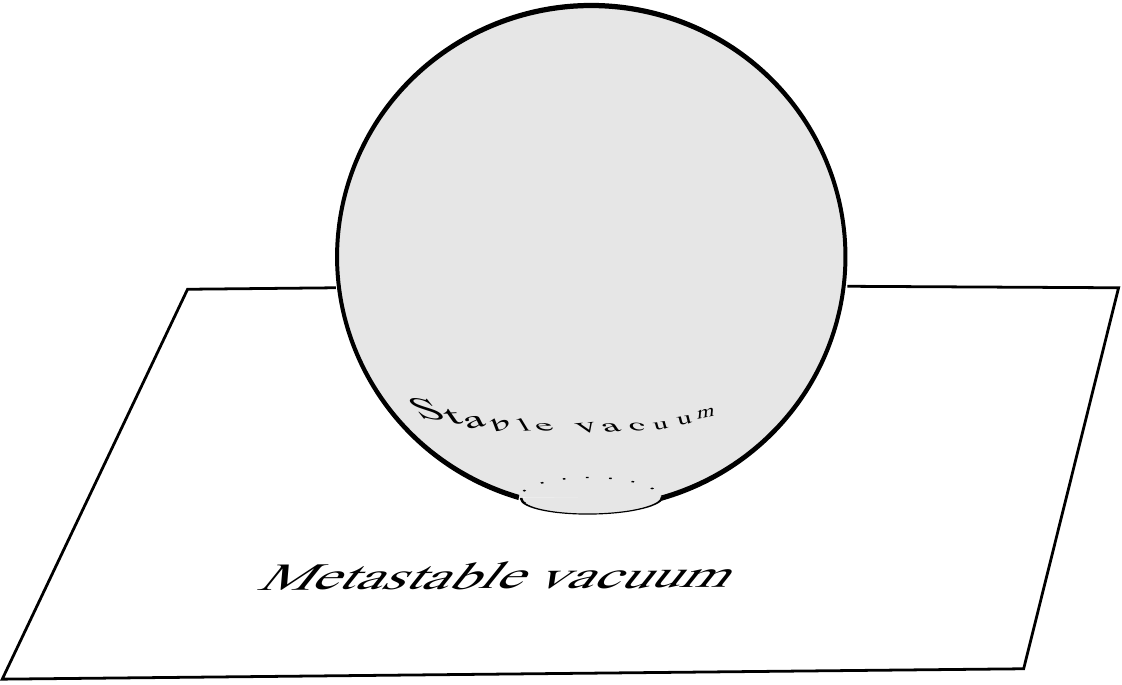}
 \caption{\small The metric of the critical droplet in presence of
 gravity.  }
\label{fig:Bubble1}
         \end{minipage}
      \end{figure}

In IQG, the discontinuity of $\CG_\low $ vanishes near the origin as
$\xi^{3/2}$ and not exponentially, as predicted by the droplet model
\cite{AlZ-unp}.  A field-theoretical explanation of the power law was
given in \cite{Zamolodchikov:2006xs}, where Langer's theory of the
condensation point was adjusted to the case of a fluctuating metric.
In the case of a flat lattice, the discontinuity across the Langer cut
is related to the Boltzmann weight of the critical droplet of the true
vacuum in the sea of the false vacuum (Fig.\ref{fig:Droplet}).  A. and
Al.  Zamolodchikov argued in \cite{Zamolodchikov:2006xs} that in
presence of gravity the droplets of the stable phase influence the
metric, so that the perimeter of the boundary of a droplet grows
slower than the square of the area (Fig.  \ref{fig:Bubble1}).  This
leads to a much faster decay of the unstable vacuum, which changes the
character of the singularity at $\xi=0$.  This phenomenon was named by
the authors of \cite{Zamolodchikov:2006xs} `critical swelling'.

The work \cite{Zamolodchikov:2006xs} gave a convincing evidence that
the Langer singularity in IFT has its counterpart in IQG, and that
both singularities are due to the presence of a weakly metastable
state.  Is it possible to extend the IFT/IQG correspondence to more
general spin systems?  In fact, the analysis of
\cite{Zamolodchikov:2006xs} applies to any droplet model in presence
of gravity.  Due to the critical swelling, the Boltzmann weight of the
critical droplet is expected to behaves as a power, $h^{ g_\low}$,
where the exponent $ g_\low$ is determined by the conformal anomaly in
the low-temperature phase $c_{\text{low}}$.  The saddle point analysis
of \cite{Zamolodchikov:2006xs} gives $g_\low =- {1\over 6}
c_{\text{low}}$ in the classical limit $c_{\text{low}}\to -\infty$,
where the fluctuations of the Liouville field can be neglected.  The
authors of \cite{Zamolodchikov:2006xs} conjectured (we will refer to
this as Z-Z conjecture) that the exact nucleation exponent $g_\low$ is
related to the central charge of the stable phase as follows,
   \be\la{nuclexp}   g_\low = {p\over p-1} \quad \text{for}
   \quad c_{\low}= 1- {6 \over p(p-1)}.  \ee
In the case of IQG, $c_{\low}=0$ and $g_\low = 3/2$, which is in
agreement with the singularity of $\CG_\low$ at $\xi=0$.  If the
conjecture \re{nuclexp} is confirmed, this would open the possibility
to study the analytical properties of the free energy of a spin
systems on a flat lattice, in particular in what concerns the Langer
and YL singularities, using its exact solution on a dynamical lattice.

The conjecture \re{nuclexp} about the nucleation rate exponent can be
tested by looking for more solvable examples among the dynamical
lattice statistical systems, which exhibit metastability and have
non-trivial massless modes in the stable vacuum.  In this paper we
argue that the $O(n)$ loop model in external magnetic field is such a
system.

Technically the main result of this paper is the explicit expression
for the specific free energy of the gravitational $O(n)$ model in the
continuum limit, which we obtain using the correspondence with a large
$N$ matrix model \cite{Kostov:1988fy}.  The main body of the text is
devoted to the interpretation of this exact solution.  In Sect.
\re{sec2} we formulate the expected analytic properties of the free
energy of the $O(n)$ loop model on a flat lattice as well as the the
non-perturbative effects associated with the existence of a
hypothetical metastable vacuum.  In Sect.  \re{section:dynamical} we
perform the same analysis of the gravitational $O(n)$ model.  In
particular, assuming the existence of a metastable state and the
conjectured nucleation exponent \re{nuclexp}, we reproduce the form of
the series expansion of the free energy of the low-temperature phase
at $h=0$.  In Sect.  \ref{section:sol} we verify that the exact
expression for the free energy possesses the expected analytic
properties, which justifies the assumption of the existence of a
weakly metastable state in the $O(n)$ loop model.  Our result verify
the Z-Z conjecture for the whole spectrum $-\infty <c<1$ of the
central charge of the matter field.  Since the paper is somewhat
technical, below we give a short summary of our results.
 
 \medskip
 
\centerline{$\diamond   \diamond \diamond    $}

 \medskip
 
\begin{figure}
         \centering
         \begin{minipage}[t]{0.4\linewidth}
            \centering
            \includegraphics[width=5cm]{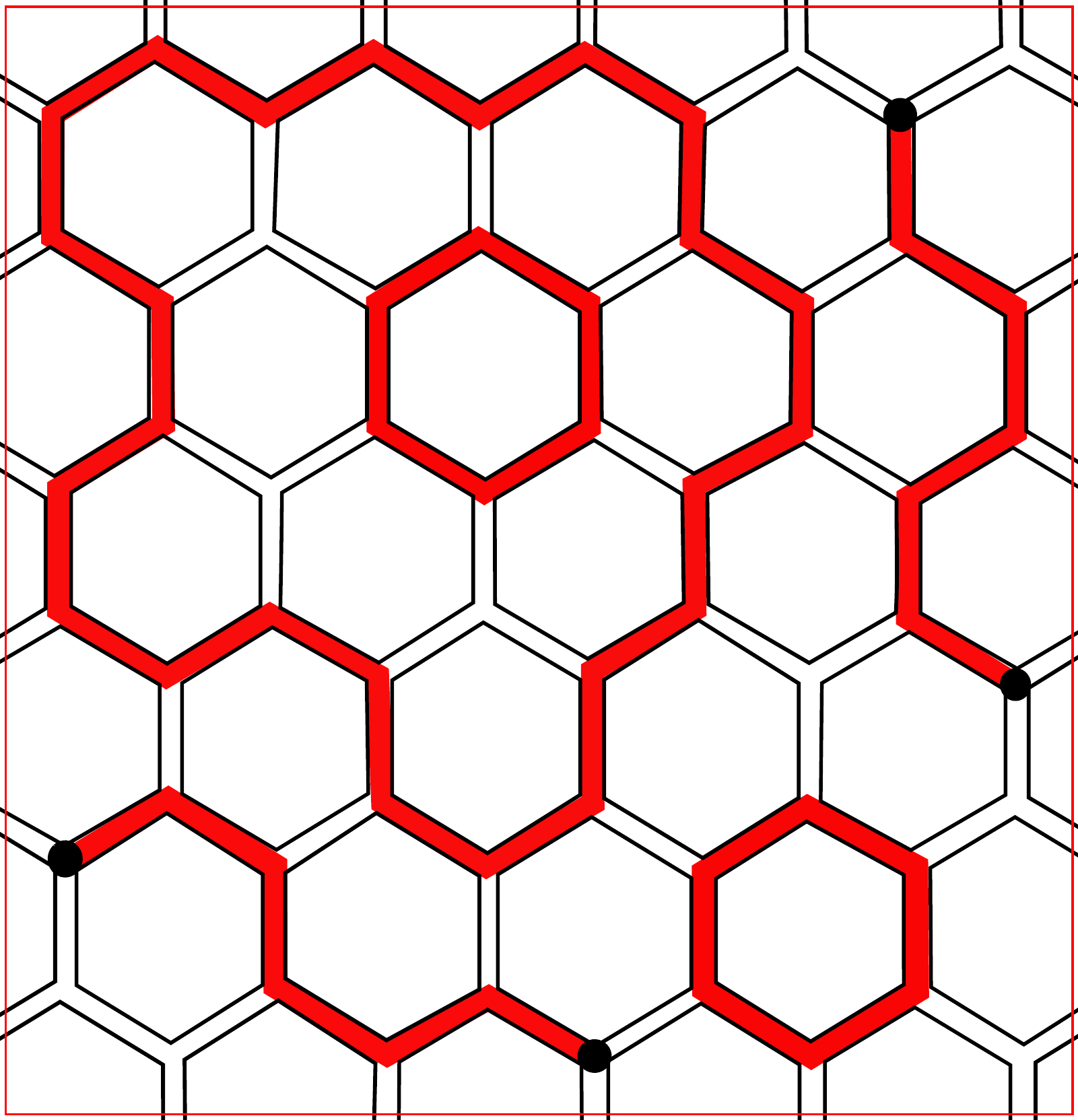}
 \caption{ \small A polymer configuration on a honeycomb lattice}
 \label{fig:Honeycomb}
         \end{minipage}%
         \hspace{2cm}%
         \begin{minipage}[t]{0.4\linewidth}
            \centering
            \includegraphics[width=6cm]{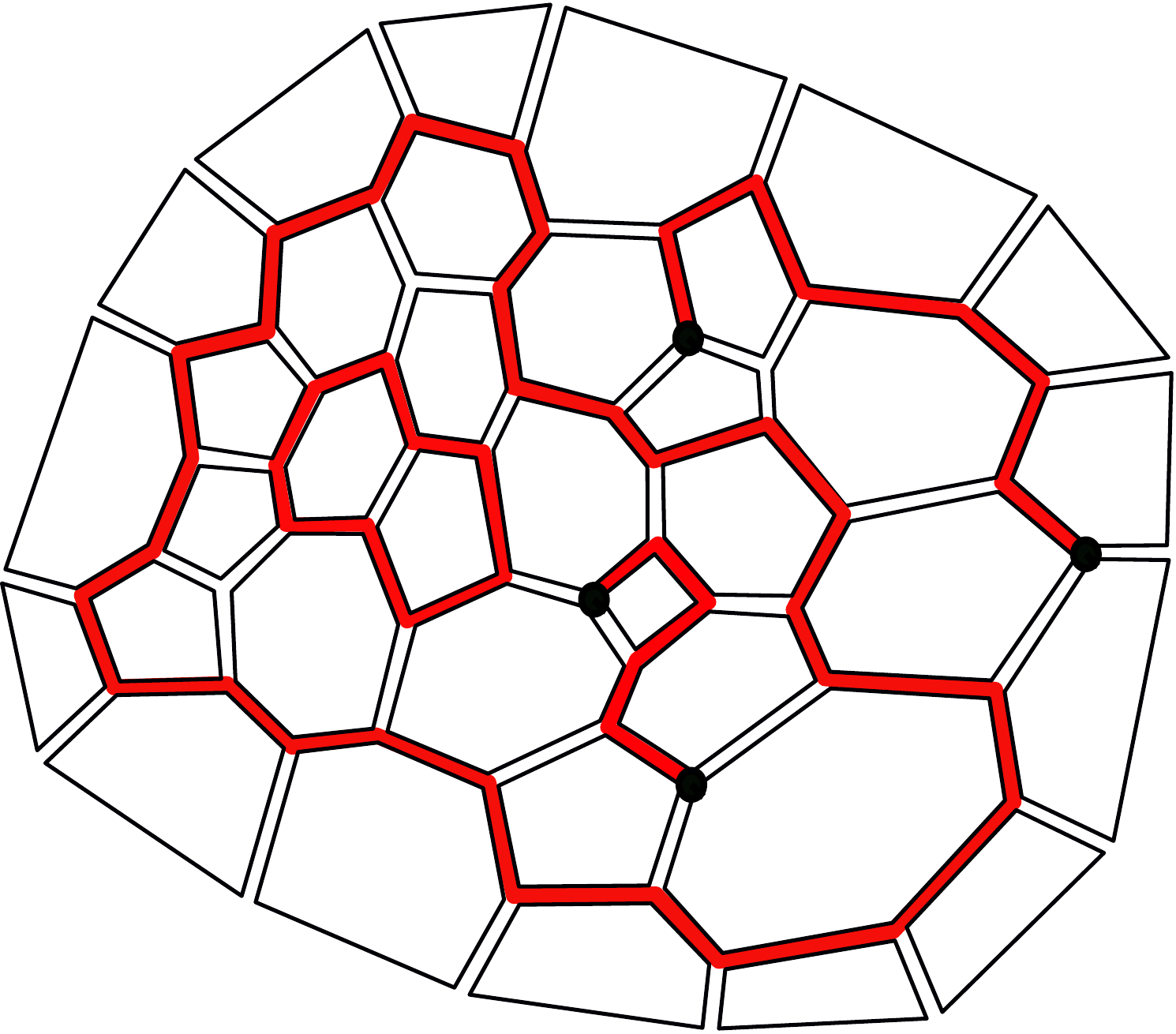}
 \caption{\small A polymer configuration on a planar graph with the
 topology of a disk}\label{fig:Disk}
         \end{minipage}
      \end{figure}

The $O(n)$ loop model \cite{Domany:1981fg,Nienhuis1982} has a
geometrical formulation in terms of nonintersecting loops with
fugacity $n$.  In presence of magnetic field $H$, the geometrical
expansion involves also open lines with fugacity $H^2$, as in the
example shown in Fig.  \ref{fig:Honeycomb}.  The temperature $T$
controls the total length of the linear polymers.  In the conformal
window $-2<n<2$, the theory has a conformal invariant critical point
at $T=T_c$ and $H=0$, described by a CFT with central charge
\cite{Nienhuis:1987aa},
\be\la{ccriticalp} c_{_{\rm crit}}= 1- {6 \over p(p+1)}, \ee
where $p$ is related to the dimensionality $n$ by
\begin{equation}
 \label{parnp} 
 n= 2\cos{\pi \over p} , \qquad p\ge 1.
\end{equation}
The critical $O(n)$ model covers the whole one-parameter family of
universality classes having massless modes in the low-temperature
phase, with a central charge $-\infty<c\le1$.

The phase diagram of the $O(n)$ loop model is similar to that of the
Ising model, which corresponds to the particular case $n=1$, or $p=3$.
For $T>T_c$ the typical linear polymer is short and the theory has a
mass gap.  Nienhuis \cite{Nienhuis1982} showed, using the mapping to
other solvable lattice models and the Coulomb gas techniques, that for
$T<T_c$ and $H=0$ the model is characterized by a massless
low-temperature (LT) phase, also known as a phase of dense loops.  The
LT phase of the $O(n)$ vector model is described by another CFT with
central charge\footnote{To avoid confusion, let us mention that this
low-temperature phase is not a Goldstone phase.  The $O(n)$ loop model
flows to a Goldstone phase having central charge $n-1$ appears when
the loops are allowed to intersect with non-zero probability
\cite{Martins:1997zn, Jacobsen:2002wu}, and in the microscopic
formulation given in \cite{Domany:1981fg,Nienhuis1982} such
intersections are strictly forbidden.  }
 \be \la{cdensep} c_{\text{low}}= 1- {6 \over p(p-1)}.
\label{cden}
 \ee

The vicinity of the critical point ($T=T_c, H=0$) is described by the
`$O(n)$ field theory', or shortly $O_n$FT, which is a perturbation of
the CFT with central charge \re{ccriticalp}.  The $O(n)$ field theory
is formally defined by the action
\be\la{ActOnFT} \CA_{{\OnFT} } = \CA_{{_{\rm crit}} }- h \int
 \s (x)\, d^2 x- t \int \varepsilon (x)\, d^2 x \, , \ee
where ${ \s}$ and $\varepsilon $ are the conformal fields generating
the perturbations respectively in the magnetic field $h\sim H$ and in
the temperature $t\sim T-T_c$.  When $p$ is an odd integer, $p= 2m-1$,
this action describes the unitary minimal conformal theory ${\cal
M}_{2m-1,2m}$ perturbed by the two primary fields
\be \la{Phihphit} \hskip 2cm \varepsilon  = \Phi_{1,3} \, , \quad \s =
\Phi_{m,m} \hskip 2cm (m= \hf (p+1))\, .  \ee
The Ising field theory corresponds to $m=2$, or $n=1$.  The thermal
flow generated by $\varepsilon $ ends, depending on the sign of the
temperature coupling, at a massive theory in the HT regime, or at a
CFT with the lower central charge \re{cdensep} in the LT regime
\cite{Nienhuis:1984wm}.  For a general $p>1$, the $O(n)$ field theory
does not exist as a local field theory, but it has nonetheless an
unambiguous definition in terms of grand canonical ensemble of linear
polymers on the lattice.

In Sect.  \ref{sec2} we speculate about the analytic properties of the
specific free energy of $\OnFT$, assuming that there exists a weakly
metastable state and Langer singularity at least in some finite
vicinity of the point $p=3$.  Since the free energy depends
analytically on the parameter $p$, this is a natural assumption,
although for $p\ne 3$ we do not see any geometrically transparent
picture of the nucleation mechanism.\footnote{ Even in the Ising case,
$p=3$, it is not clear how to characterize the unstable phase in terms
of the gas of dense self-avoiding loops and lines.}
We give a heuristic argument about the
form of the Langer singularity  for $p\ne 3$.   The
non-perturbative corrections   should be of the
form
  \be \la{standardroplet1}
   F_{\text{nonpert}}
   \sim  f\ \exp(-\text{const}/ f ), \qquad
  f \sim |h|^{ 1/(1-\Delta_\s^{\text{low}})},
  \ee
where $\Delta_\s^{\text{low}} $ 
is the conformal dimension of 
the $O(n)$ spin field $\s $ in the low-temperature  phase.

As in the case of IFT, the specific free energy is determined by two
different scaling functions in the HT and in the LT regimes:\footnote{
We find more convenient to introduce two different, although related
by a phase factor, dimensionless variables, $\xi$ for the HT regime
and $\z$ for the LT regime.  }
  \be F(t,h) = \begin{cases}\ \ t^{p+1\over 2} \ G_\high(\xi), \qquad
  \xi= t^{-\kappa_0} h\quad & \text{if } \quad t>0, \\
      &\\
     (-t)^{p+1\over 2} \ G_\low(\z), \quad \z= (-t)^{-\kappa_0} h\quad
     & \text{if} \quad t<0,
\end{cases}
  \ee
  where the power ${p+1\over 2}$ and the exponent
\be\la{defnuo} \kappa_0\equiv {1- \Delta_{\s } \over 1
- \Delta_{\varepsilon }} = {(3p+1)(5p+3) \over 32 p}\, \ee
are determined by the conformal weights $\Delta_\varepsilon =\bar
\Delta_\varepsilon $ and $\Delta_\s=\bar \Delta_\s$ respectively of
$\varepsilon $ and $\s $ at the critical point.  The function
$G_\high$ has a couple of Yang-Lee branch points on the imaginary axis
and is analytic at the origin, while the function $G_\low$ is expected
to have an essential singularity at $\z=0$.  We argue that the domain
of analyticity of $G_\low(\z)$ is the wedge $|\arg(\z)|<\pi(1-
\Delta_\s^\low)$.  The scaling functions $G_\high$ and $G_\low$ are
analytically related at infinity and can be expressed through a third
scaling function $\Phi(\eta)$, where $ \eta = h^{ -1/\kappa _0} \, t$
is the dimensionless temperature coupling.

In Sect.  \ref{section:dynamical} we formulate the the analytic
properties of the free energy of $O(n)$ quantum gravity, or shortly
$\OnQG$, which describes the continuum limit of the $O(n)$ model on a
dynamical lattice.  The geometrical expansion in this case involves
non-intersecting loops and lines on planar graphs as the example in
Fig.  \ref{fig:Disk}.  $\OnQG$ is formally defined as a perturbation
of Liouville gravity with matter central charge \re{ccriticalp} by the
action
\begin{eqnarray}
\label{actgIsOn}
\delta {\cal A}_{\OnQG} = -t \, \int \varepsilon (x) \,
e^{2\a_\varepsilon \phi (x) } d^2 x - h \, \int \s (x) \, e^{2 \a_\s
\phi(x)} d^2 x,
\end{eqnarray}
which generalizes \re{actgIsi}.  The vacuum energy of $\OnQG$ must be
of the form
\be \CF(t,h) = \begin{cases}\ \ t^{p} \ \CG_\high(\xi), \qquad \xi=
t^{-\kappa} h\quad & \text{if } \quad t>0, \\
      &\\
(-t)^{p}\ \CG_\low(\z), \quad \z= (-t)^{-\kappa} h\quad & \text{if}
\quad t<0,
\end{cases}
  \ee
where $\kappa$ is the ratio of the Liouville dressing exponents
associated with the two perturbing operators,
  \be
  \kappa ={\a _\s \over \a_\e }= {3p+1\over 4}.
  \ee
 Again, the functions $\CG_\high$ and $\CG_\low$ are related
 analytically at infinity and can be expressed in terms of a third
 scaling function $\Phi(\eta)$, where $\eta= h^{-1/\kappa} t$ is the
 dimensionless temperature.  This determines the form of the expansion
 at infinity
 \be \la{expbigxi} \CG_\high(\xi)=\sum_{j=0}^\infty \Phi_j\ \xi^{1-j
 \a_\e \over \a_\s }, \qquad \CG_\low(\z)=\sum_{j=0}^\infty (-1)^j\ \Phi_j
 \ \z^{1-j\a_\e \over \a_\s }.  \ee
At the origin the function $\CG_\high$ is analytic, while $\CG_\low$
is expected, according to \cite{Zamolodchikov:2006xs}, to have a
power-like singularity.  The leading term in the expansion is $f \sim
\z^{1/\a_\s ^\low}$, where $\a_\s ^\low$ is the Liouville exponent
associated with the magnetic operator in the low-temperature phase.
The Z-Z conjecture concerns the subleading term, which should be $\sim
f^{\, g_\low}$, with $g_\low$ given by \re{nuclexp}.  We develop
further the arguments of \cite{Zamolodchikov:2006xs} by considering
the contributions of droplets-within-droplets configurations.  In this
way we were able to predict not only the first two terms, but the form
of the whole series expansion the scaling function $\CG_\low$,
\be \begin{aligned}\CG_\low(\zeta) &=\tilde \CG_0+ C_1\, f + C_2 \,
{f}^{\, g_\low} \, \big(1+ C_3 \, {f^{\, g_{\low} } \over \z } +\dots
\big) \\
 &=\tilde \CG_0+ \sum_{n\ge 0} \tilde \CG_n \
 \z^{{4(p-1)\over 3p- 1} + { 4 n\over 3p- 1}}.
\end{aligned}
 \la{ZZexp1} 
   \ee

In Sect.  \ref{section:sol} we compare the speculations of sections
\ref{sec2} and \ref{section:dynamical} about the behavior of the
$O(n)$ free energy with the exact solution, obtained for the
microscopic realization of $\OnQG$ as the $O(n)$ loop model on a
dynamical lattice.  The derivation, based on the correspondence with
the $O(n)$ matrix model, is given in Appendix \ref{sec:MatrixH}.  The
specific free energy $ \CF(t, h)$ was found in a parametric form,
   \be \la{equ519a} \begin{aligned} \CF(t, h) =& - { y^p\over 2} + t \,
   {y^{ p-1} \over 2(1-{1\over p}) } - h^2\, {y^{-(p+1) /2}\over
   1+{1\over p} }\, ,\\
 h^2\ =&\ y^{(3p-1) /2 }( y- \, t ).
\end{aligned}
 \ee
For $p=3$ the equation for the $O(n)$ free energy coincides, up to a
shift by a term $\sim t^3$, with Boulatov-Kazakov solution of IQG, eq.
\re{FreeIQG}.
  
From the solution \re{equ519a} we find the scaling functions
$\CG_\high$ and $\CG_\low$.  They represent two different branches of
the same meromorphic function $\CG$, which is defined as the analytic
continuation of $\CG_\high(\xi) $ under the YL cuts.  The function
$\CG(\xi)$ has the symmetries
  \be \la{symmetryRS}
  \CG(\xi) = \CG(-\xi) =\overline{ \CG(\bar\xi)},
  \ee
which are inherited by those of the function $\CG_\high$ and which
preserve by construction the HT sheet.  Starting from the HT sheet and
taking different paths, one can achieve (for general $p$) four
different copies of the LT sheet, related by the symmetries
\re{symmetryRS}.  The HT and the LT sheets are connected by a {\it
finite} number of auxiliary sheets.  The extended analyticity
assumption, which is fulfilled here, means that there is no additional
singularities on the connecting sheets.

We computed the series expansions of $\CG_\low$ and $\CG_\high$ at
$\z=0$ and at $\z=\infty$.  The expansion of $\CG_\low$ at the origin
is indeed of the form \re{ZZexp1}, which confirms the Z-Z conjecture
in its stronger form used in Sect.  \ref{section:dynamical} as well as
the existence of Langer singularity at the origin associated with the
presence of a metastable state.  If the gravitational field is
`switched off', the power-like singularity at $\z=0$ should turn into
an essential singularity associated with this metastable state.  Let
us stress that we expect Langer-type singularity only in the $O(n)$
loop model which has a representation in terms of a gas of
non-intersecting polymers.  The conventional $O(n)$ model should not
have such a singularity.

To summarize, in this work we argue that Langer's singularity 
in presence of small magnetic field, originally observed in the Ising
model,  is in fact a general feature of
the $O(n)$ loop models.  We are quite confident in our conclusions when $0<n<2$, while in the interval $-2<n<0$, where the loop gas does not have 
statistical interpretation, the situation is less clear.  Our exact expression for the free energy  of the gravitational $O(n)$ model presents a
strong evidence about the existence of a metastable state in the
low-temperature phase and confirms the Z-Z conjecture about the decay
rate of the metastable vacuum in presence of gravity
\cite{Zamolodchikov:2006xs} in the case when the stable phase has
non-zero central charge.  We found that the scaling function
$\CG(\xi)$ for the free energy of the $O(n)$ loop model on a dynamical
lattice has a pair of Yang-Lee singularities on the high-temperature
sheet, a Langer-type singularity of the expected form on the
low-temperature sheet, and no additional singularities on the
connecting sheets.  In this sense the scaling function obeys the
`extended analyticity' assumption of Fonseca and Zamolodchikov
\cite{Fonseca:2001dc}.  However, it is not clear if this property will
remain true for the theory on a flat lattice.  This is a very
important question which deserves further study.

\section{The vacuum energy of the $O(n)$ field theory}
\label{sec2}

The analytic properties of the free energy of of $O_nFT$, except
Langer singularity, follow immediately from the known scaling behavior
near the critical points, where the theory becomes conformal
invariant, and the assumption that on the first sheet there are no
other singularities than the critical points.  These predictions are
in accordance with the numerical results for the cases $n=1$
\cite{Fonseca:2001dc,Mangazeev:2009rm} and $n=0$
\cite{PhysRevB.35.3657}.  In contrast, as we already discussed, we do
not have direct arguments in favor of the presence of Langer
singularity.  This is a conjecture which is justified by the exact
solution of the model coupled to gravity.

\subsection{Continuous transitions and effective field theory on a
regular lattice}

The local fluctuating variable in the $O(n)$ loop model
\cite{Domany:1981fg,Nienhuis1982,nienhuis1983critical} is an
$n$-component vector $\vec S(r)$ with unit norm, associated with the
vertices $r $ of the honeycomb lattice, and interacting in an
isotropic way along the links $<\!\!r, r'\!\!\!>$.  In the presence of
a constant magnetic field $\vec H$, the energy of a spin configuration
is defined by the ``geometric'' hamiltonian
\be \la{defHH} \CH[\{\vec S\}] =- \sum_{<r, r'> } \log\( T + \vec
S(r)\cdot \vec S(r')\) -\sum_{r }\log\(1+ \vec S(r)\cdot \vec H\).
\ee

The logarithmic form of the nearest neighbor interaction leads to a
simple graphical expansion which generalizes that of the Ising model
and can be mapped to a solvable vertex model.  The partition function
is defined as the trace (the integral over all spin configurations)
  \be \CZ_{_{O(n)}}(T; \Gamma) =\Tr \ e^{- \CH[\{\vec S\}]}, \ee
where by the $O(n)$ symmetry $\Tr ( S_a S_b )= \delta_{a,b}$ and $\Tr
(S_a) = \Tr (S_a S_b S_c)=0$.  Expanding the integrand as a sum of
monomials and using the rules above one can write the high temperature
expansion of the partition function as the grand canonical ensemble of
non-intersecting polymers of variable length.  There are two kinds of
polymers: loops with activity $n$ and open lines with activity $ H^2$,
where $H= |\vec H|$.  Apart of the activity, the Boltzmann weight of
each polymer is equal to $T^{-L}$, where $L$ is the length, defined as
the number of links covered by the polymer.  The high temperature
expansion of the partition function is a triple series in $n$, $1/T$
and $ H^2$,
\begin{equation}\label{looprON}
 \CZ_{_{O(n)}}(T, \vec H)=\sum_{\text{polymers }}\; (1/T)^{L_{\rm
 tot}}\, n^{\#\text{loops}} \, H^{2 \#\, \text{open lines}}.
\end{equation}
The temperature coupling $T$ controls the total length $L_{\rm tot}$
of the polymers (the number of links covered by loops or open lines).
An example of a polymer configuration is given in Fig.
\ref{fig:Honeycomb}.  The geometrical expansion \re{looprON} of the
$O(n)$ model allows to consider the number of flavors $n$ as a
continuous parameter.

When $\vec H=0$, the $O(n)$ loop model is known to have a continuous
transition in the window $-2\le n\le 2$, where $n$ can be parametrized
by eq.  \re{parnp}.  The phase diagram of the loop gas on the
honeycomb lattice was first established by Nienhuis
\cite{Nienhuis1982}.  At the critical temperature $T_{\text{crit}} =
2\cos \frac{\pi(p-1)}{4p}$ the loop gas model is solvable and
described by a CFT with central charge \re{ccriticalp}.  This point is
usually referred as the `dilute' phase.  For $T>T_{\text{crit}}$, the
theory has a mass gap.  The low-temperature, or `dense', phase
$T<T_{\text{crit}}$ represents a massless flow toward an attractive
fixed point at $T_{\text{low}}= 2\sin \frac{\pi(p-1)}{4p}$.  At this
point the theory is again solvable and described by a CFT with smaller
central charge $c_\text{low}$ defined by \re{cdensep}.

The spectrum of the degenerate primary fields in the critical and in
the low-temperature CFTs is given respectively by
\be\la{defcd}
\Delta_{r,s}= 
{(r(p+1) - s p)^2 - 1\over 4 p(p+1)}, \qquad
\Delta_{r,s}^{\text{low}} = 
{(r p - s (p-1))^2 - 1\over 4 p(p-1)}\, , \ee
where $r$ and $s$ are integers.  The thermal operator $\varepsilon $,
which counts the total length of the polymers in the expansion
\re{looprON}, can be identified with the degenerate field
$\Phi_{1,3}$.  Added to the action, it generates a mass of the loops.
The spin operator $\s$ coupled to the magnetic field corresponds to a
degenerate primary field only if $p$ is an odd integer,
\be p= 2m-1.  \ee
However, it is convenient to extend the Kac parametrization \re{defcd}
also for non-integer $r$ and $s$.  Then the magnetic field can be
identified with the field of conformal dimensions $\Delta_\s =\bar
\Delta _\s = \Delta_{m,m}$, with $m ={1\over 2}(p+1)$.  The magnetic
operator is the first of an infinite series of $L$-leg operators
$\Phi_{L m , L m }\ ( L=1,2,\dots)$ \cite{Duplantier:1986zza}.
Explicitly, the conformal weights of the thermal and spin operators
are given by
\be\la{thdims} \Delta_\varepsilon = \Delta_{1,3} = {p-1\over p+1},
\quad \Delta_\s = \Delta_{{p+1\over 2} , {p+1\over 2} } ={1\over 16} \
{(p+3) (p-1)\over p(p+1)}
 .
  \ee
Both operators are relevant.  The $O(n)$ field theory, which describes
the continuum limit of the Hamiltonian \re{defHH}, is formally defined
by the action \re{ActOnFT}, where
 \be t\sim T-T_\text{crit}, \quad h\sim |\vec H|.  \ee
parametrize the vicinity of the critical point of the lattice model.
When $n=1$, or $p=3$, this action coincides with the action
\re{ActIsingFT} for the Ising field theory.  Depending on the sign of
the temperature, the perturbation by the thermal operator $\varepsilon$
generates a flow toward the massive or to the dense phase.\\

As we already mentioned in Sect.  \ref{sec:Introduction}, the
low-temperature phase of the $O(n)$ loop model defined by the
Hamiltonian \re{defHH} does not contain Goldstone modes.  This is in
contrast with the standard $O(n)$ model \cite{PhysRevLett.20.589,
kurtze1979yang}, which is defined by the Hamiltonian
\begin{equation}\la{Stanley_H}
\CH[\{\vec S\}] =- {1\over T} \sum_{<r, r'>} \vec
S(r)\cdot \vec S(r') -\sum_{r}\vec S(r)\cdot \vec H.
\end{equation}

In the case of the Ising model ($n=1$), the Hamiltonians \re{defHH}
and \re{Stanley_H} are equivalent, up to a redefinition of the two
couplings.  For general $n\in[-2,2]$, the two Hamiltonians are
supposed to share the same universal properties in the high
temperature phase and at the transition point, while the low
temperature phases of the two models are believed to be different
\cite{Jacobsen:2002wu}.  For $T<T_c$, the $O(n)$ symmetry of the model
with Hamiltonian \re{Stanley_H} is spontaneously broken to $O(n-1)$,
which leads to a massless Goldstone phase with central charge
$\tilde{c}_\text{low}=n-1$.  (Since the model is non-unitary, the
Mermin-Wagner theorem does not apply here.)  In contrast, the
low-temperature phase of the model with Hamiltonian \re{defHH}
exhibits unbroken $O(n)$ symmetry.
    
In terms of the geometrical expansion, the Goldstone phase appears
when the loops are allowed to intersect.  The Hamiltonian \re{defHH}
of the $O(n)$ loop model is designed in such a way that the
intersections are strictly forbidden.  This is why the Goldstone modes
do not appear.  It is argued in \cite{Jacobsen:2002wu} that an
arbitrarily small perturbation that allows intersections would cause a
crossover to the generic Goldstone phase.

 \subsection{High-temperature regime}
 
In the high-temperature phase of the $O(n)$ model, the deformation
\re{ActOnFT} with $t >0, h=0$ describes, from short to long distance
scales, the flow to a massive theory.  For finite $t$, the specific
free energy, or the vacuum energy density of the $O(n)$ field
theory,\footnote{If the theory is defined on a cylinder of radius $R$,
the specific free energy per site is determined by the leading
asymptotics of the energy of ground state $E_0(R)$, when the radius
tends to infinity, $F=\lim_{R\to\infty} E_0(R)/R$.} scales as
 \be \la{ashzero} F(t, 0)   \sim  t^{1-\Delta_\varepsilon }=
    t^{p+1\over 2}  
 \qquad ( h=0, \ t>0) .  \ee
 %
For finite $h$ and $t$, the specific free energy must have the scaling
form
\be \la{scalingRHT} F (t, h) =   t^{p+1\over 2} G_\high (\xi), \ee
where the scaling function $G_\high $  depends on the 
dimensionless variable
\be\la{defnuob} \xi = t^{-\kappa_0} h, \quad \kappa_0= {1- \Delta_{h }
\over 1- \Delta_{t}} = { (3p+1)(5p+3) \over 32 p}\, .  \ee
When $p$ is an odd integer, as in the case of the IFT, one should take
the derivative of \re{scalingRHT} in $p$,
\be 
\la{defpoddF}
F(t,h) \sim 
t^m\log t + t^m \hat G_{\high}(\xi),
\qquad p= 2m-1= \text{odd integer}.  \ee

Since for a finite positive $t$ the theory has a mass gap, the scaling
function should be analytic at $\xi=0$.  The high temperature
expansion \re{looprON} involves only even powers of the magnetic
field, which means that $G_\high $ is an even function of $\xi$ in the
vicinity of $\xi=0$,
 \be \la{asymHT} G_\high (\xi)  =
       G_0+G_2 \xi^2 + G_ 4\xi^4 + \dots  \quad \text{for}\ \ \xi\to 0.
 \ee
On the other hand, the large $h$ finite and $t=0$ the theory is again
massive with correlation length determined by the conformal weight
$\Delta_\s$.  Hence the asymptotic behavior of the scaling function at
infinity is
 \be \la{asymHTinf} G_\high (\xi) \sim
    \xi^{ 1/(1-\Delta _\s )} +\cdots \quad  \text{ for}\ \ \xi\to\infty.
 \ee

 The function $G_\high (\xi)$ can be analytically continued for
 complex values of $\xi$.  Obviously the scaling function has branch
 points at some finite $\xi$, otherwise the two asymptotics
 \re{asymHT} and \re{asymHTinf} would be incompatible.  As argued in
 \cite{kurtze1979yang}, the general $O(n)$ model exhibits in the HR
 regime the same analytical properties than the Ising model: the zeros
 of the partition are situated along the imaginary axis, and an
 infinite strip centered around the real axis is free of zeros.  Then,
 the free energy would have two symmetric branch cuts on the imaginary
 axis, extending from $ \pm i \xi_c$ to $\pm i\infty$, as is shown in
 Fig.  \ref{fig:HTcuts}.  The branch points correspond to the two
 Yang-Lee edges, and in the vicinity of $\xi = \pm i\xi_c$ the $O(n)$
 field theory is expected to be described by the minimal CFT
 $\mathcal{M}_{2,5}$ with a central charge $c^\YL = -22/5$
 \cite{Cardy:1985yy}.  The Yang-Lee CFT has only one relevant operator
 $\Phi_{1,2}$ with the conformal dimension $\Delta^\YL_{1,2} = - 1/5$.
 The dimension of the corresponding coupling constant is
 $1-\Delta_{1,2}^\YL=6/5$ and one expects, as in
 \cite{Fonseca:2001dc}, that near the YL branch points $G_\high (\xi)$
 behaves as
\be G_\high (\xi) = g_A(\xi) + g_B (\xi)\, \(\xi_c^2 +\xi^2\) ^{5/6}+
\dots , \ee
with $g_A$ and $g_B$ being regular functions of $\xi$.

 \subsection{Low-temperature regime}

In the LT regime ($t<0$)  the theory flows toward a
CFT with the central charge $c_\low$ given in
\re{cdensep} \cite{Nienhuis:1984wm, Fendley:1993wq}.  
The perturbation of  the low-temperature CFT 
at  $t\to -\infty, h=0$   is driven by  the operators
$\varepsilon$ and $\s$ with conformal weights  
  \be \Delta^{\text{low}}_\varepsilon = \Delta_{3,1}^\low = {p+1\over
  p-1}, \quad \Delta^{\text{low}}_\s = \Delta_{{p-1\over 2}, {p-1\over
  2}}^\low= {1\over 16} \ {(p-3) (p+1)\over p(p-1)} \ee
and coupled respectively to $-1/t$ and $h$.  The energy becomes an
irrelevant operator, while the magnetic field remains relevant.  A
finite perturbation at $t\to -\infty$ with this operator leads to a
massive theory with specific free energy
  \be \la{LTF} F(t, h) \sim h^{1/(1-\Delta^{\text{low}}_\s )}
 \qquad ( t\to -\infty).
 \ee
For the Ising model $(p=3)$ the low-temperature phase contains no
degrees of freedom, $c_\text{low}=0$ and $ \Delta_\s ^{\text{low}}=0$.
The free energy is $F\sim h$ and the magnetization
\be
\bar \s  = -\p F/\p h
\la{defMh}
\ee
is a constant.  The dimension of the magnetic operator becomes
negative for $p<3$.  Accordingly, the magnetization $\bar \s $
vanishes at $h\to 0$ for $p>3$ and diverges as $h^{\Delta_\s
^{\text{low}}/(1-\Delta_\s ^{\text{low}})}$ for $p<3$.  In particular,
for $p=2$, which corresponds to the case of the dense polymers, the
specific free energy diverges as $F\sim h^{32/35}$.  This scaling
behavior has been tested and confirmed numerically by Saleur
\cite{PhysRevB.35.3657}.  In the opposite limit, $h\to\infty$, the
free energy is again given by \re{asymHTinf}.  On the other hand, for
finite negative $t$ and $h=0$ the free energy behaves as
\be \la{scalingLTho} F(t, h) \sim (-1/t)^{1/( 1-\Delta^\low_\varepsilon)}
=
(-t)^{p+1\over 2} 
\qquad ( h=0, \, t<0).
 \ee
 For finite values of $h$ and $t$ the vacuum energy is of the form
\be \la{scalingRLT} F(t, h) =(-t)^{p+1\over 2} \, G_{\low} (\z), \ee
where $G_\low$ is the scaling function in the low-temperature phase
and $\z$ is defined as
\be\la{defnuzb} \z =(- t)^{-\kappa _0} h \, .  \ee
At small $\z$ the scaling function $G_\low$ has asymptotic behavior
     \be \la{asymlowG} G_\low(\z) \sim \tilde G_0 + \tilde G_1 \, \z^{
     1/(1-\Delta _\s ^{\text{low}})} + \dots \quad \text{for} \ \z\to
     0 , \ee
  while at infinity $ G_\low\sim \z^{ 1/(1-\Delta _\s )}$.  By
  construction the function $G_\low(\z)$ is real for $\z>0$.  The
  constant term in \re{asymlowG} is in general different than the one
  in \re{asymHT}.  Since the susceptibility and the magnetization are
  positive, we have $\tilde G_0>0$ and $\tilde G_1<0$.  The relation
  between $ G_\low$ and $ G_\high$ will be obtained in Sect.
  \ref{sec:Phi}.

 \subsection{Langer singularity}
 
In the Ising model ($p=3$) the scaling function $G_\low$ is expanded
in the integer powers of $\z$, but the power series is asymptotic.
The analytical continuation of the free energy from positive $\z$ has
an exponentially small purely imaginary discontinuity on the negative
axis, which is explained by the presence of a metastable state for
small negative $\z$.  In the language of QFT, one could say that the
theory contains a global resonance state with complex energy, the
`false vacuum'.  The imaginary part of the specific free energy can
then be interpreted as the decay rate per unit time and volume of this
resonance state
\cite{Kobzarev:1974cp}\cite{PhysRevD.15.2929}\cite{Callan:1977pt}.
Langer \cite{langer1967theory, Langer:1969aa} developed a systematic
approach, based on the earlier droplet models, to evaluate the
discontinuity $\d F= 2i \Im F$ of the free energy on the negative
axis.  The latter decays exponentially for small magnetic field, $\d
F\sim \exp({-{\rm const}/|\z|})$.

By construction, the free energy and the correlations in the $O(n)$
loop model are analytic in $p$.  If the perturbative expansion of the
free energy at $p=3$ is asymptotic, it will be asymptotic also in the
vicinity of $p=3$.  Therefore it is very likely that a metastable
state and Langer type singularity should exist also for $p\ne 3$.
This conjecture will be later confirmed by the exact solution on
dynamical lattice.  Unfortunately it is not clear how to characterize
geometrically the true and the false vacuum within the loop gas
representation, even in the Ising case $p=3$.

In the Ising model the energy gap between the stable and the
metastable vacuum is evaluated by analytically continuing the free
energy to negative values of $h$.  For general $p$ the leading
behavior of the free energy for small $h$ is no more linear in $\z$,
 \be \la{pertFLT} F(t,h) = (-t)^{p+1\over 2} \( \tilde G_0 + \tilde
 G_1\z^{2\nu}+\dots\) , \quad 2 \nu\defeq 1/( 1- \Delta_\s ^\low), \ee
and the metastable state should be reached by analytical continuation
to the rays $\arg (\z) = \pm \pi/2\nu$ in the complex $h$-plane.  Here
the free energy is again real and the gap
\be 
\begin{aligned}
f(t, \z) \equiv  F ( t, e^{ \pm  i \pi /2 \nu}h)
-  F(t, h)  
\end{aligned}
\la{frentrue} 
\ee
is positive.  Langer's theory \cite{langer1967theory} predicts a
nnon-perturbative, exponentially small in $f$, imaginary part of $F$,
equal to the decay rate $\Gamma$ (per unit volume and unit time) of
the metastable vacuum,
\be
F  (t, e^{ i\pi/2 \nu} h)-F  (t, e^{- i\pi/2 \nu} h)\sim i \Gamma.
\ee

To evaluate the leading exponential factor, assume that initially the
two-dimensional plane is filled with the metastable vacuum with
specific free energy $F+f$ and imagine a hypothetical stochastic
process which allows the statistical system to develop in time.  The
decay of the metastable vacuum will then occur through spontaneous
formation of droplets of the stable phase in the ``sea'' of metastable
phase.  The Boltzmann weight of a droplet of area $A$ and perimeter
$L$ is
 \be W_{A,L} \sim e^{ A  f  - L \sigma}
, \ee
 where $\s $ is the surface tension at the boundary of the droplet.
 The surface tension scales as $\sigma= \sigma_0 (-t)^{p+1\over 4}$,
 where $\s_0$ is a numerical coefficient.  Assuming that the drops are
 approximatively round, $A = L^2/ 4\pi $, the integral in $L$ is
 saturated by the saddle point $L_{s.p.} = 2\pi \sigma/f $.  The
 saddle point value gives the size of the `critical droplet', which
 can trigger a decay of the metastable state.  For droplets larger
 than the critical droplet the bulk energy starts to decrease faster
 than the surface energy grows and the droplet energetically favorable
 to grow.  The leading factor in the decay rate $\Gamma $ is the
 Boltzmann weight of the critical droplet.  A more refined calculation
 \cite{Voloshin:1985id}, taking into account the fluctuations of the
 form of the droplets, gives also the pre-exponential factor,
 \be
\la{standardroplet2}
 \begin{aligned}
 \Gamma
 \sim 
      {f} \ e^{ -  \pi \sigma^2/  f}. 
     \end{aligned} \ee
 Therefore, with the assumptions we made, the scaling function
 $G_\low(\z)$ for the LT regime develops an essential singularity
 $\sim |z|^{2\nu} e^{- |\z|^{2\nu}}$ at the origin.  The function
 $G_\low(\z)$ is analytic in the $\z$-plane with a cut starting at
 $\z=0$, which can be placed along one of the rays $\arg (\z) = \pm
 \pi/2\nu$, where the discontinuity of the leading perturbative term
 vanishes, or along the negative axis, which will be our choice in the
 following.

\subsection{The scaling function $\Phi(\eta)$}

\la{sec:Phi} The functions $G_\high (\xi)$ and $G_\low(\z)$, although
defined for two different phases, must be analytically connected in
the regime where the magnetic field is sufficiently large compared to
the temperature.  In this regime it is useful to introduce the
dimensionless temperature
\be\la{etay} \eta =   h^{ -1/\kappa _0} \, t,
\ee
with  $\kappa _0$ was defined in \re{defnuo}, and a new scaling
function $\Phi(\eta)$ by
\be F(t, h) = h^{1/(1-\Delta_\s )}\, \Phi(\eta).  \ee
This new variable is suited to study the analytical property of the
free energy in terms of the temperature.  In particular, it allows the
description of the neighborhood of the critical point where $t\sim0$.
In the regime $t>0$ and $h>0$, the variable $\eta$ is related to $\xi$
by
\be\la{etaxiHT} \eta = \xi^{-1/\kappa _0} \ee
and the new scaling function is expressed in terms of $G_\high (\xi)$
as
\be\la{PhiGHT} \Phi(\eta) = \eta^{{1\over 1-\Delta_\varepsilon }} \,
G_\high (1/ \eta^{\kappa _0}), \qquad (\eta>0).  \ee
Similarly, in the regime $t<0$ and $h>0$
\be\la{etaxiLT} \eta = -\,  \z^{-1/\kappa _0} \ee
and  
\be\la{PhiGLT} \Phi(\eta) = (-\eta)^{{1\over 1-\Delta_\varepsilon }} \,
G_\low(1/(-\eta)^{\kappa _0}), \qquad (\eta<0).  \ee
(When $p$ 
is an odd  integer, the above relations are slightly modified.
 They are obtained by taking  the derivative in $p$ and using
  the definition \re{defpoddF}.)

\begin{SCfigure}
  \centering
  \includegraphics[width=0.37\textwidth]%
    {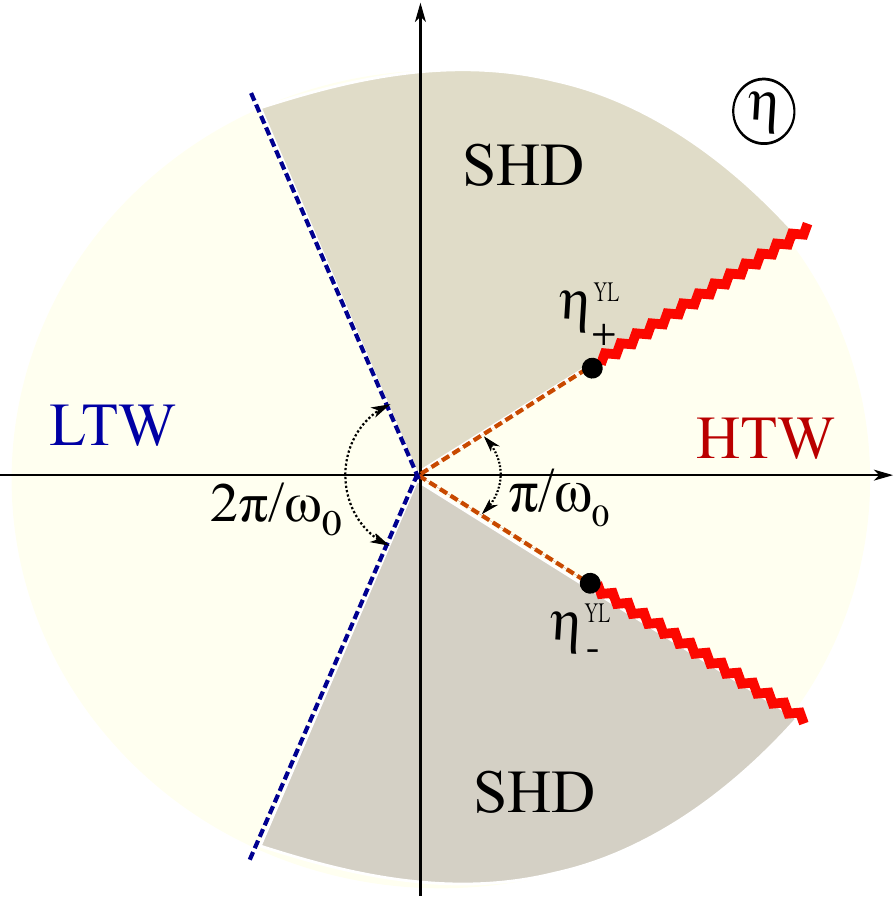}
\hspace{1cm} \caption{\small Principal sheet of the $\eta$-plane for
the scaling function $\Phi(\eta)$.  The LT wedge $
|\arg(-\eta)|<{2\pi/ \kappa_0}$ is the image of the principal sheet of
the scaling function $G_\low(\z)$.  The HT wedge $ |\arg(\eta)|<{\pi/
2\kappa_0}$ is the image of the right half plane of the principal
sheet of $G_\high(\xi)$.  The two sectors SHD separating the LT and
the HT wedges are the `shadow domain' where the analytic properties of
$\Phi(\eta)$ are not known.  In \cite{Fonseca:2001dc} it was
conjectured (for $p=3$) that the function $\Phi(\eta)$ is analytic
also in the shadow domain.}
\label{fig:etaplane}
\end{SCfigure}

Since for $h\ne 0$ the theory has a mass gap, the free energy $F( t,
h)$ should be analytic in the temperature $t$ in some finite strip
around the real axis.  Therefore the scaling function $\Phi(\eta)$ is
represented by a power series
\be\la{Tayloreta} 
\Phi(\eta) = \Phi_0+ \Phi_1\, \eta+
\Phi_2\eta^2 + \dots 
\ee
with a finite radius of convergence near $\eta\sim0$.  According to
\re{PhiGHT} and \re{PhiGLT}, the coefficients $\Phi_k$ of the series
\re{Tayloreta} determine the large expansion of the scaling functions
$G_\high (\xi)$ and $G_\low(\z)$ at infinity,
\be 
\begin{aligned}
G_\high(\xi)&=&\xi ^{ {1\over 1-\Delta_\s }} \Phi(\xi^{- 1/\kappa_0})
\quad = \sum_{j=0}^\infty \Phi_j\ \xi^{1-j(1-\Delta_\varepsilon )\over
1-\Delta_\s }, \qquad\\
G_\low(\z) &=&\z ^{ {1\over 1-\Delta_\s }} \Phi(-\z^{- 1/\kappa_0})  =\
\sum_{j=0}^\infty (-1)^j \Phi_j \ \zeta^{1-j(1-\Delta_\varepsilon )\over
1-\Delta_\s }.  
\end{aligned}
\la{GPhi}
\ee
In the vicinity of $\xi=\infty$, the function $G_\low (\z)$ can be
obtained from $G_\high(\xi)$ by analytical continuation $\xi\to
\z=e^{\pm i \pi \kappa_0 } \xi$.

The domains of analyticity of $G_\high$ and $G^\low$ are mapped to
different (and non-overlapping for $p\ge 2$) domains on the principal
sheet of the scaling function $\Phi(\eta)$.  The map $\xi\to\eta$
sends the positive real axis of the $\xi$-plane to the positive real
axis of the $\eta$-plane.  The right $\xi$-half-plane, where
$G_\high(\xi)$ is analytic, is mapped to the high temperature wedge
(HTW) of the $\eta$-plane, defined as
 \be
\text{HTW:}
\quad
 -{\pi\over 2\kappa _0} <\arg \eta < {\pi\over 2\kappa
_0}. 
 \ee
In particular, the LY branch points are mapped to \be \eta _\pm ^\YL=
e^{\pm i\pi /2\kappa _0} \ \eta_c, \quad \eta_c =\xi_c ^{1/\kappa _0}.
\ee 
 One can choose the branch cuts starting at the YL edges to go to infinity
 along the rays $\arg\eta = \pm\pi/2\kappa _0$.  (Note that the cuts
 in the $\eta$-plane are not the images of the cuts of the
 $\xi$-plane.)

On the other hand, the map $\z\to\eta$ sends the positive real axis in
the $\z$-plane to the negative real axis of the $\eta$-plane.  The
principal sheet of $G^\low$ in $\z$-plane is mapped to the low
temperature wedge (LTW),
\be
\text{LTW:}\quad -{\pi \over \kappa _0} <\arg(- \eta) < {\pi \over
\kappa _0} .  \ee
In the $\eta$-plane the Langer branch cut is resolved and the Langer
singularity is sent to infinity.  The two edges of the cut are mapped
to the two rays $\arg(-\eta)=\pm { \pi /\kappa _0}$.  The principal
sheet of the function $\Phi(\eta)$ is shown in Fig.
\ref{fig:etaplane}.

As in the Ising model, the function $\Phi(\eta)$ is expected to be
analytic in the sectors HTW and LTW of the $\eta$-plane.  The two
sectors separating HTW and LTW were called in \cite{Fonseca:2001dc}
`shadow domain'.
 If the extended analyticity, claimed in \cite{Fonseca:2001dc}, holds
 for all $p\ge 2$, which means that $\Phi(\eta)$ is analytic also in
 the shadow domain, then the scaling function can be reconstructed
 unambiguously from its discontinuities along the two Yang-Lee cuts on
 the principal sheet of the $\eta$-plane.

\section{The vacuum energy of the gravitational $O(n)$ model}
\la{section:dynamical}

In this section we obtain the expected analytical properties of the
free energy of the $O(n)$ model coupled to 2d gravity.  We adjust the
arguments of the previous section to the situation when gravity is
switched on, making substantial use of the Z-Z conjecture about the
effect of the critical droplet.

\subsection{Continuous transitions and effective field theory on a
dynamical lattice}
\label{subsec:contflat}

The $O(n)$ model on a dynamical lattice has the same continuous
transitions as the theory on a flat lattice.  In the continuum limit
the sum over the planar graphs with a given topology is replaced by a
functional integral with respect to the Riemann metric $g_{ab}(x, \bar
x)$ on a variety with the same topology.  Here, we will restrict
ourselves to the case of the sphere and the disc, where the metric can
be put in the form $g_{ab}(x, \bar x) = \delta_{ab} \, e^{2\phi(x,
\bar x)}$ by a coordinate transformation.  On the sphere, the
integration measure with respect to the scale factor $\phi$, known as
the Liouville field, is defined by the action \cite{Polyakov:1981rd}
\be \la{Laction} \CA_{\text{Liouv}} = {g\over 4\pi} \int (\partial
\phi)^{2} d^2 x +\mu \int e^{2 \phi }d^{2}x  \, ,
\label{LAc}
\ee
 and  the asymptotics at infinity
\be \phi(x,\bar{x})\simeq -Q\log(x\bar{x}), \qquad Q= g+1.  \ee
The Liouville coupling constant $g$ is determined by the critical
fluctuations of the matter field and the `cosmological constant' $\mu$
controls the global area of the sphere
 \be \la{defaria} A=  \int e^{2\phi }\, d^2 x .  \ee
When the topology is that of a disk, the integral \re{Laction} is
restricted to the upper half plane, with an extra boundary term $\mb
\int e^\phi dx$.  The boundary coupling constant $\mb$ controls the
length of the boundary of the disk,
 \be L=  \int \, e^\phi\,dx .  \ee
The Liouville action \re{Laction} leads to a CFT with central
charge
\be \la{cliouv} c_{\ \text{Liouv}} = 1+ 6(g+1)^2/g.  \ee
Besides the Liouville field there are also Fadeev-Popov ghosts with
total central charge $c_{\text{ghosts}}=-26$ (see the standard reviews
\cite{Ginsparg:1993is} and \cite{DiFrancesco:1993nw}).  Due to the
general covariance, the total conformal anomaly in the theory of
Liouville gravity must vanish, $c_{\ \text{Liouv}}
+c_{\text{matter}}+c_{\text{ghosts}}=0$.  When the matter field is the
$O(n)$ CFT with the central charge \re{ccriticalp}, this constraint
determines the Liouville coupling constant $g$ in \re{Laction} as
\be g= (p+1)/p .  \ee
The observables in Liouville gravity are the integrated local
densities
\begin{eqnarray} 
\la{defOh} { \mathcal{ O}_\Delta } \sim \int \Phi_\Delta \, e^{2\alpha
\phi}d^2x\, ,
\end{eqnarray}
where $\Phi_\Delta $ represents a scalar ($\Delta=\bar \Delta $)
matter field, and the vertex operator $e^{2\a\phi}$ is the Liouville
dressing factor, which takes into account the fluctuations of the
metric.  The Liouville dressing factor completes the conformal
dimensions of the matter field to $(1,1)$.  Only such, marginal,
operators are allowed in the theory coupled to gravity.  The balance
of the conformal dimensions gives a quadratic relation between the
Liouville dressing charge $\a$ and the conformal weight of the matter
field, known as the KPZ relation
\cite{Knizhnik:1988ak,David:1988hj,distler1989conformal}
\be 
\label{relaD}   { \alpha ( g+1 -\alpha )/ 4  g} =1- \Delta \, .  \ee
 The `physical' solution of the quadratic equation \re{relaD} is the
 smaller one.

In the Kac parametrization \re{defcd} of theconformal weight
$\Delta=\Delta_{r, s}$ of the matter field, the physical solution of
the quadratic equation \re{relaD} is given by
\begin{eqnarray}
\label{scaldims} \alpha_{rs}(g) = {2p+1 - |r(p+1)-s\, p |\over 2\, p}.
\end{eqnarray}
 In particular, the operators
 \be \CO_\varepsilon = \int \Phi_\varepsilon \, e^{2\a_\e \phi} \, ,
 \quad \CO_\s = \int \Phi_{h}\, e^{2\a_\s \phi} \ee
represent the Liouville-dressed thermal and magnetic operators with
dimensions \re{thdims}, and the corresponding Liouville exponents are
\be \a_\e \equiv \a_{1,3}= {1\over p}, \quad \a_\s \equiv
\a_{{p+1\over 2},{p+1\over 2}} = { 3p+1 \over 4 p}.  \ee
The $O(n)$ gravity is formally defined by the action (omitting the
ghost term)
  \be\la{aacag} \CA _{\OnQG} =\CA_{\rm crit} + \CA _{\text{Liouv} }- t \ 
  \CO_{\varepsilon} - h  \  \CO_{\s }.  \ee

Since all operators in Liouville gravity are marginal, the dimensions
of the matter fields cannot be extracted from the coordinate
dependence of correlation functions of the theory.  In fact, the
correlation functions in Liouville gravity do not depend on the
coordinates at all.  The dimension of the matter component of an
operator $\CO=\Phi\ e^{2\a\phi}$ is determined by the response to a
rescaling of the metric, that is, to a translation of the Liouville
field.  As a consequence of the KPZ relation \re{relaD}, the exponent
$\a$ is positive if the matter field $\Phi$ is relevant and negative
when the matter field is irrelevant.

Let $\CZ(t,h, \mu)$ be the partition function of Liouville gravity on
the sphere.  The role of the RG time in Liouville gravity is played by
the zero mode of the Liouville field, which is equal to the logarithm
of the square root of the area \re{defaria}.  Therefore sometimes it
is instructive to consider instead the partition function on the
sphere for fixed area, which is given by the inverse Laplace transform
with respect to the cosmological constant,
\be \CZ _A(t, h) =\int_{\uparrow} {d\mu\over 2\pi i} \ e^{A\mu}\
\CZ(t, h, \mu) . \ee
Here the contour of integration goes on the right of all singularities
of $\CZ(t, h, \mu) $.  In the limit of infinite area the partition
function behaves as
\be \CZ_A(t, h) \sim A^{- 7/2}\ e^{ - A\, \CF(t, h) }.  \ee
The exponent is by definition the specific free energy for the $O(n)$
quantum gravity.

\subsection{Analytic properties of the specific free energy in the
high-temperature regime} \la{subsec:GHT}

In the previous section, the scaling behavior of the specific free
energy of the $O_n$FT near the critical points was determined by the
dimensions of the two coupling constants, $t$ and $h$.  After coupling
to gravity, the dimensions of the coupling constants reflect the
response of the corresponding operators to translations of the zero
mode of the Liouville field.  With the convention that the
cosmological constant $\mu$ has dimension 1, the dimension of the
coupling constant for the operator \re{defOh} is $\a$.
 
To establish the scaling properties of the free energy in $O_n$QG, one
can essentially repeats the arguments of the previous section,
replacing the dimensions $1-\Delta$ of the coupling constants by the
Liouville exponents $\a$.  In the high-temperature regime, $t>0$, the
specific free energy of the gravitating $O(n)$ field should be of the
form

\be \la{scalingHTd} \CF(t, h) = t^{1/\a_\e } \ \CG_\high  (\xi), \ee 
and the scaling function $\CG_\high(\xi) $ depends on the dimensionless
strength of the magnetic field, defined as
\be \la{defxih} \xi= t^{ - \kappa }\, h\, , \quad \kappa = { \a_\s 
\over \a_\e }
= { 3p+ 1  \over 4}.
\ee
 For finite positive $t$, the theory has a mass gap and the scaling
 function must be even and analytic near $\xi=0$:
 \be \la{asymHTb} \CG_\high (\xi) =\CG_0+ \CG_2 \xi^2 + \CG_ 4\xi^4 +
 \dots , \qquad |\xi|<\xi_c .  \ee

The free energy is expected to have two symmetric Yang-Lee branch cuts
on the imaginary axis\footnote{The YL singularity in the IQG was first
discovered in \cite{Staudacher:1989fy}}, extending from $\xi^\YL_\pm
=\pm i \xi_c$ to $\pm i\infty$, with a value $\xi_c$ different from
the flat lattice model one.  Near the YL singularity, the theory is
that of a Liouville gravity with $c_\text{matter}=-22/5$, or $g= 5/2$.
The magnetic field is coupled to the only primary field $\Phi_{1,2}$,
which have Liouville dressing exponent $\a_{1,2}= 3/2$.  In such
non-unitary theory, the operator that determines the behavior at large
distances (here large $\phi$) is the most relevant operator
$\CO_{1,2}$, while the perturbation with $\delta\xi \cdot \CO_{1,1}$,
$\delta\xi \equiv \xi - \xi^\YL_\pm$, leads to a `correlation area'
$A_c \sim (\delta \xi )^{- \a_{1,2}/\a_{1,1}} =(\delta\xi)^{-3/2}$.
Therefore the specific free energy must behave for $\xi \approx
\xi_\pm ^\YL$ as
 \be \la{YLpred} \CG _\high (\xi ) \sim \text{reg.}\ + \( \xi^2 +\xi
 _c^2 \)^{3/2} +\cdots\ee
where ``reg'' denotes regular terms in $\delta\xi$.

\subsection{The specific free energy in the low-temperature regime and
the Langer singularity}
\la{subsec:GTL}

As in the case of rigid geometry, when $t\to -\infty $, the theory
flows to the low-temperature CFT in which the coupling constant of the
Liouville field is
\be \la{defglow} g_\low = p/(p-1).  \ee
The thermal operator becomes irrelevant, but the magnetic operator
remains relevant,
\be \a_\e ^{\text{low}} \equiv \a_{3,1}^\low= - {1\over p-1}, \quad
\a_\s ^{\text{low}}\equiv \a^\low_{\frac{p-1}{2},\frac{p-1}{2}} = {
3p-1\over 4 (p-1)}.  \ee
A finite perturbation at $t\to -\infty$ with the magnetic operator
leads to a massive theory with specific free energy
 \be 
 \la{CFsmalh}
 \CF(h) \sim h^{1/\a^{\text{low}}_\s } = h^{4(p-1)\over 3p-1},\qquad
 t\to -\infty.  \ee

For finite values of $h$ and $t<0$ the free energy has  the form
\be \la{defGlow} \CF(t, h) =(- t)^{p} \ \CG_\low (\z).  \ee
where $\CG_\low$ is the scaling function in the low-temperature phase
(analytically continued from a real positive magnetic field) and $\z$
is the dimensionless variable
\be
\la{defzeta}
\la{defnuz} \z =(- t)^{-\kappa } h \, .  \ee
According to \re{CFsmalh}, the leading behavior at small $\z$ of the
scaling function is
 \be \la{asymxiLT} \CG_\low(\z)=  
 \tilde \CG_0 + \tilde\CG_1 \,
  \z^{ 1/ \a_\s ^{\text{low}}} +\text{subleading terms}, \qquad
 \z\to 0 .  \ee
Since the magnetization $\bar \s= \p \CF/\p h$ is  positive, 
the coefficient $ \tilde\CG_1$ must be negative.

Assuming that there is a metastable state, its energy is obtained by
analytic continuation from the positive axis of the $\z$-plane to the
ray $\arg\z= \pi \a_\s ^{\text{low}}$, where the leading free energy
is again real.  The energy gap between the metastable and the true
vacuum is now
\be
\la{defff}
f(t, \z)  \equiv \CF(t, e^{i\pi  \a_\s ^{\text{low}}}h )-  \CF(t,  h) 
=- 2 \tilde\CG_1\,
(-t)^p \, \z^{1/  \a_\s^\low} >0.
\ee
 
Let us sketch the semiclassical argument presented by A. and Al.
Zamolodchikov in \cite{Zamolodchikov:2006xs}.  Assume that in the
beginning we have an infinite, globally flat space filled with the
metastable phase.  The corresponding classical solution for the metric
is $\phi = 0$.  Assume also that there is a stable phase with
negative, compared to that of the metastable phase, energy density $-f
$ and that the decay of the metastable phase goes through formation of
droplets of the stable phase.

\def\CF{{\cal F}} Let us calculate the energy of a configuration with
one circular droplet of the stable phase which covers the circle
$|x|<r$.  Then the total energy of the droplet is
  \be \CF_{\text{droplet}}= -f \int _{|x|<r}d^2 x \ e^{2\phi(x)}
  +\sigma \int_{|x|=r} e^{\phi(x) }dx , \ee
where $\sigma$ is the surface tension at the boundary of the droplet.
The metric in the presence of the droplet is given by the extremum of
the Liouville action \re{Laction} with $g$ and $\mu$ replaced by
 \be
 \begin{aligned}
 g&\quad \to\quad g_{_{{\text{low}}}}= p/(p-1), \\
  \\
       \mu& \quad \to\quad \begin{cases}-f  & \text{ if } \ |x|<r, \\
     0 & \text{if }\ |x|>r.
\end{cases}
\end{aligned}
  \ee
 The corresponding classical solution is obtained by sewing the
  solution of the Liouville equation 
  \be \la{Ggliouv}  g_\low\  \p_x\p_{\bar x} \phi (x,\bar x) = -f \,  
  e^{2\phi (x,\bar x)}   \ee
 inside the circle with the flat solution $\phi=0$ outside.  Since the
 cosmological constant is {\it negative} inside the droplet, the
 general solution of \re{Ggliouv} with radial symmetry describes a
 metric with constant {\it positive} curvature:
  \be e^{2\phi(x)}= {4 R^2 a^2 \over (1+a^2 |x|^2)^2}\, .
   \ee
This is the metric of a sphere with radius $R$, which is related to
$f$ by
  \be R^2 = {g_{_{{\text{low}}}}\over 4 \pi f}.  \ee
  %
This solution should be sewed with the solution $\phi=0$ for $|x|>r$.
This gives the condition $1+ a^2 r^2 = 2 a R$, which has two solutions
for $a$,
  \be a_\pm = {R \pm \sqrt{R^2-r^2}\over r^2}, \ee
with an obvious geometrical meaning.  Imagine that the sphere of
radius $R$ is cut into two disks along a circle with radius $r$.  Then
the two disks correspond to the two possible solutions for the
parameter $a$.  The smaller disk ($a=a_-$) resembles a cap and the
larger disk ($a=a_+$) resembles a bubble.  The saddle point for the
integral in $r$ and $R$ gives for the energy of the critical droplet
$E_c = g_\low \log ( 1+ \pi \sigma^2 /g_\low f )$.  For $g_\low f\gg
\sigma^2$ the unstable direction at the double saddle point is mostly
along the ``perimeter growth" direction $r$, just as in the case of
the flat geometry.  In the opposite limit $g_\low f\ll \sigma^2$, the
growth of the droplet is mostly in the ``inflation" direction $R$.
The decay rate
  \be
  \Gamma\sim e^{-E_c}= 
  \( 1+ {\pi \sigma^2  \over g_\low f}\)^{-g_\low}
  \ee
matches in the limit $g_\low f\gg \sigma ^2$ (strong metastability)
the standard droplet model result \re{standardroplet2}, while in the
limit $g_\low f\ll \s^2$ (weak metastability) it gives a power law,
$\Gamma \sim \left( g_\low f/\pi \s^2 \right)^{g_\low}$.

To summarize, in the limit of weak metastability the critical droplet
looks like a bubble connected to the rest of the surface by a thin
neck, as shown in Fig. \ref{fig:Bubble1}.  In this case the decay rate
behaves as $f^{g_\low}$ where $g_\low$ is defined by \re{defglow}.
This phenomenon was called in \cite{Zamolodchikov:2006xs} `critical
swelling'.

Although the above argument is valid only in the quasi-classical limit
$ c_\low\to-\infty$, where $g_\low = - c_\low/6$, it was argued in
\cite{Zamolodchikov:2006xs} that the power law persists for finite
$c_{\text{low}}$, and that $g_\low$ defined by \re{defglow} gives the
exact nucleation exponent.  This is supported by the following
heuristic argument.  In the limit of weak metastability the classical
solution for the metric inside the critical droplet is that of an
almost complete sphere.  One can speculate that in the quantum case
the metric within the critical droplet starts to fluctuate, but is
still connected to the rest of the surface by a thin `neck'.  If this
is true, it is easy to evaluate the exact contribution of the droplet
to the free energy.  In Liouville gravity this is the partition
function on a sphere with a puncture and a cosmological constant $-
f$, which is known \cite{Knizhnik:1988ak} to be $\sim f ^{g_\low }$.
This argument is however impeded by the negative sign of the
cosmological constant.

 \begin{figure}[h]
         \centering
         \begin{minipage}[t]{0.4\linewidth}
            \centering
            \includegraphics[width=4.0cm]{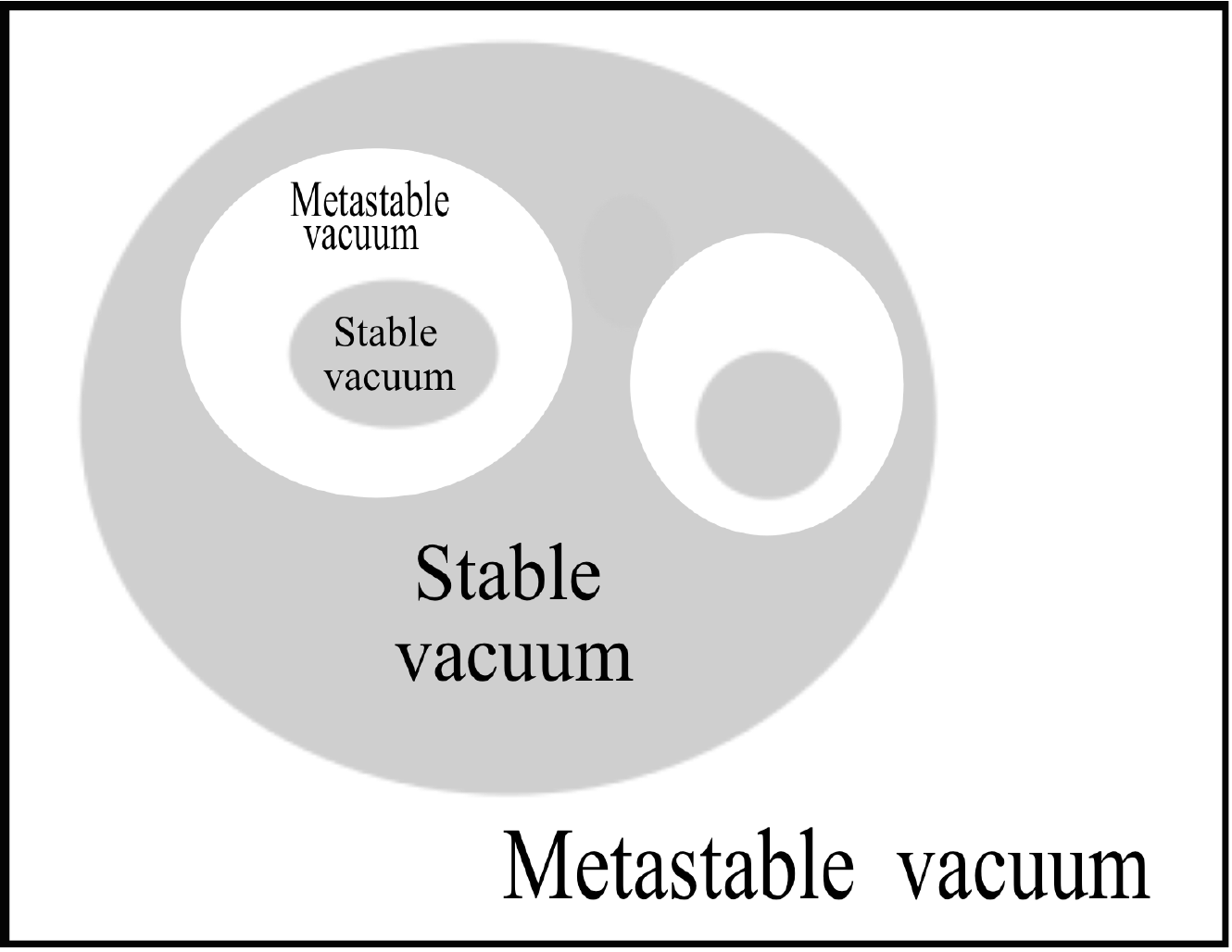}
 \caption{ \small A nested droplets configuration.  }
 \label{fig:Droplets}
         \end{minipage}%
         \hspace{2cm}%
         \begin{minipage}[t]{0.4\linewidth}
            \centering
            \includegraphics[width=5cm]{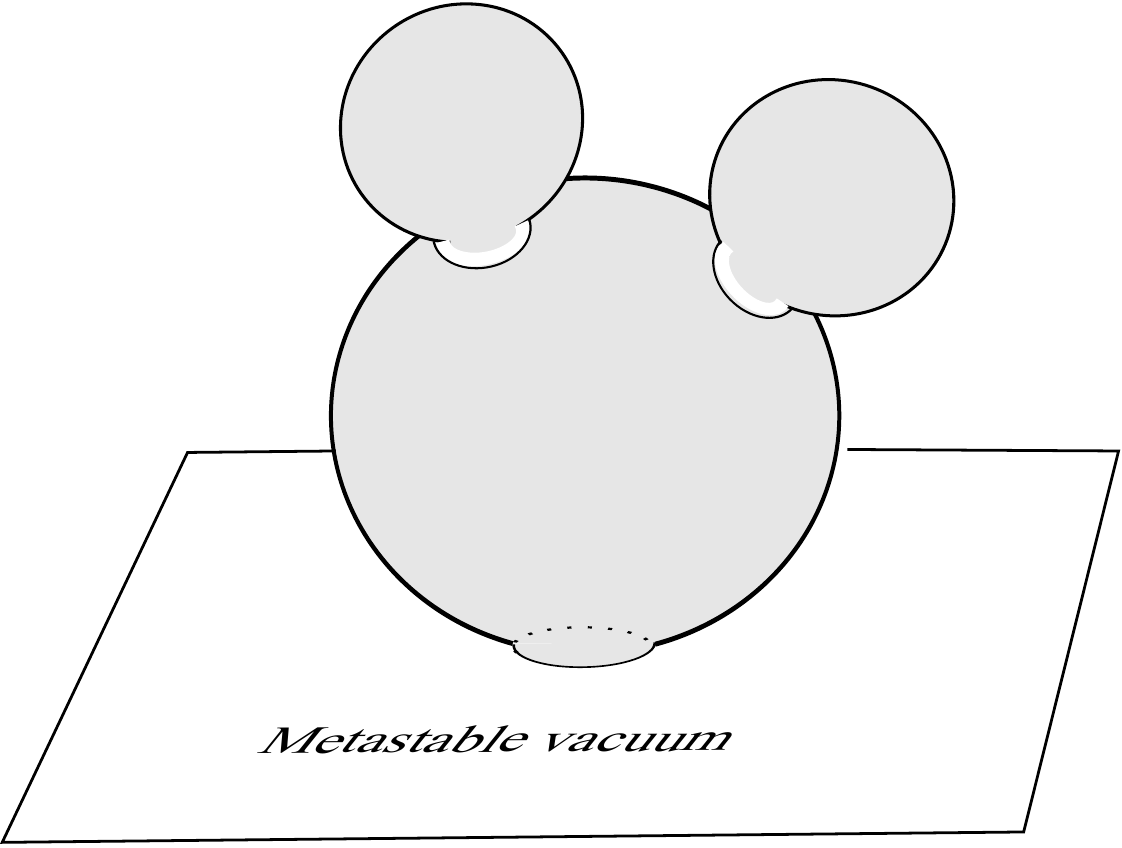}
	     \caption{\small The metric of the nested droplets
	     configuration in presence of gravity.  }
\label{fig:Bubbles}
         \end{minipage}
      \end{figure}

The Z-Z conjecture formulated above gives a prediction for the
subleading term in the small $\z$ expansion of $\CG_\low$, provided
the leading term   is known.  From  \re{defff} we find \be
\begin{aligned}
\CG_\low&= \CG_0\, + C_1 \, f + C_2 \, f^{g_\low}+\dots \\
&=\tilde \CG_0 + \tilde \CG_1\ \z^{4(p-1)\over 3p- 1} + \tilde \CG_2 \
\z^{4p\over 3p- 1}+\dots  ,
\end{aligned}
\ee
with $\tilde \CG <0$.  We can develop further the arguments of
\cite{Zamolodchikov:2006xs} by taking into account the
droplet-inside-droplet configurations as the one shown in Fig.
\ref{fig:Droplets}.  The solution of the Liouville equation for such a
nested configuration describes a metric that resembles the surface of
a cactus (without the spines), depicted in Fig.  \ref{fig:Bubbles}.
It consists of several bubbles of the stable phase connected by thin
throats of the metastable phase.  The contribution of $n$ droplets
within a droplet factorizes into the partition function of a sphere
with $n+1$ punctures and $n$ partition functions on a sphere with one
puncture.  The total power of $|\CF |$ is
 \be f^{g_\low - n} \( f^{g_\low}\)^{n} = \z ^{ {4 p\over
 3p-1} + {4n \over 3p-1}},\quad n=1, 2, \dots.  \ee
In this way the Z-Z conjecture gives not only the subleading term, but
also the complete small $\zeta$ expansion of the scaling function
$\CG_\low$:
 \be \la{ZZexp}
  \CG_\low(\zeta) = \tilde \CG_0+ 
  \sum_{n\ge 0} \tilde \CG_{n+1} \ \z^{{4(p-1)\over
 3p- 1} + { 4 n\over 3p- 1}} .  \ee
 The position of the Langer cut is at $\arg \z=\pi \a_\s^\low = \pi
 {3p-1\over 4(p-1)}$.  When $n>1$ the Langer singularity appears
 before reaching the negative axis, while for $0<n<1$ it appears after
 crossing the negative axis.

\subsection{The scaling function $\Phi(\eta)$}
\la{sect:Phieta}

To study the free energy for a large magnetic field, or equivalently
at a small temperature $t$, we introduce the scaling function
$\Phi(\eta)$ defined by
 \be \la{defPhifb}  \CF(t,h)= h^{1/ \a_\s } \Phi(\eta),
 \quad  \la{defPhif} \eta= h^{-1/\kappa }\ t\, .
  \ee
Depending on the sign of $t$, the new scaling function $\Phi(\eta)$ is
related to $\CG_\high (\xi)$ or $\CG_\low(\z) $,
\be\la{PhiGhGl}
 \Phi(\eta) =
 \begin{cases}
\eta^{{1/ \a_\e }} \, \CG_\high (1/ \eta^{\kappa }) &\quad \text{if }\
t>0\ \text{and} \ h>0, \\
(-\eta)^{{1/\a_\e }} \, \CG_\low(1/(-\eta)^{\kappa }) & \quad \text{if
}\ t<0\ \text{and} \ h>0.
\end{cases}
   \ee

As in the flat case, for a given value $h\ne 0$ of the magnetic field,
the matter theory has a mass gap and $\CF( t, h)$ should be analytic
in $t$.  Hence, the scaling function $\Phi(\eta)$ should be analytic
in a finite strip containing the real axis.  Near $\eta=0$, it is
represented by the power series
\be\la{Tayloretad} \Phi(\eta) = \Phi_0+ \Phi_1\, \eta+\Phi_2\eta^2 + \dots \ee
with finite radius of convergence.  According to \re{PhiGhGl}, the
series expansion of $\Phi(\eta)$ near the origin determines the
expansion of $\CG_\high(\xi)$ and $\CG_\low(\z)$ at infinity:
\be \la{largexiexp} \CG_\high(\xi)=\sum_{j=0}^\infty \Phi_j\ \xi^{1-j
\a_\e \over \a_\s }, \qquad \CG_\low(\z)=\sum_{j=0}^\infty (-1)^j\ \Phi_j
\ \z^{1-j\a_\e \over \a_\s }.  \ee

\begin{SCfigure}
  \centering
  \includegraphics[width=0.36\textwidth]%
    {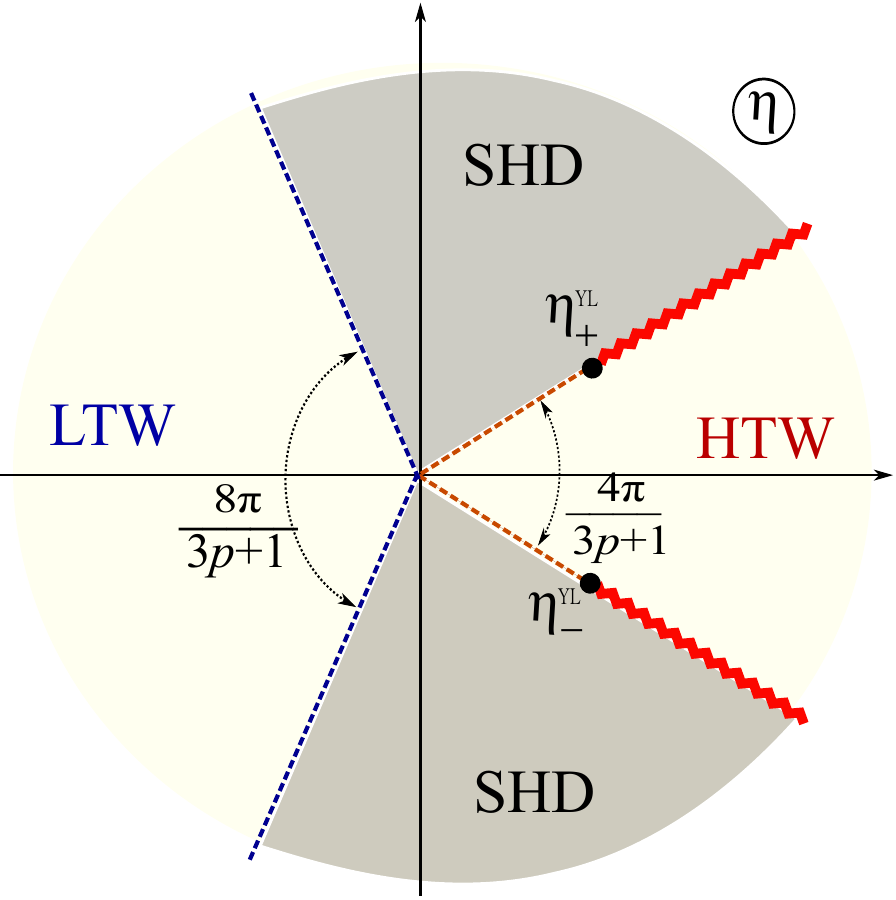}
\hspace{1cm} \caption{\small \small Principal sheet of the
$\eta$-plane for the scaling function $\Phi(\eta)$ in presence of
gravity.  The LT wedge $ |\arg(-\eta)| < {4 \pi \over 3p+1} $ is the
image of the domain of the analyticity of the scaling function
$\CG_\low(\z)$.  The HT wedge $ |\arg(\eta)| < {2\pi\over 3p+1} $ is
the image of the right half plane of the principal sheet of
$\CG_\high(\xi)$.  }
\label{fig:etaplaned}
\end{SCfigure}

The principal sheet of $\Phi(\eta)$ can be split into four sectors as
in Fig.  \ref{fig:etaplaned}.  The right half-plane $\Re\xi>0, $ where
$\CG_\high(\xi) $ is analytic, is mapped to the the HT wedge on the
principal sheet of $\Phi(\eta)$, defined as
\be \text{HTW:}\qquad  |\arg \eta | < {\pi\over
2\kappa } =  {2\pi \over 3p+1}  . \ee
The two Yang-Lee branch points  in the $\xi$ plane 
correspond to  two simple branch points of $\Phi(\eta)$ at 
\be \la{YLbps} \eta ^\YL_\pm = e^{\pm i\pi /2\kappa } \eta_c, \qquad
\eta_c = \xi_c ^{-1/\kappa }.  \ee
To exploit the analyticity of $\Phi(\eta)$ at sufficiently small
$\eta$ we place the two YL cuts such that they go to infinity along
the rays $\arg(\eta)=\pm {\pi/2\kappa }$.  Furthermore, the $\z$-plane
cut along the negative axis, where $\CG_\low(\z)$ is analytic, maps
onto the LT wedge on the principal sheet of $\Phi(\eta)$ defined as
\be \text{LTW:}\qquad  |\arg(- \eta) |
<{ \pi    \over
 \kappa }={ 4\pi\over  3p+1 } 
  .  \ee
The two edges of the Langer cut in the $\z$-plane are
mapped to the two rays $\arg(- \eta)=\pm {\pi/\kappa}$.

Coupling to gravity, or taking the average with respect to geometries,
cannot add new singularities to the free energy.  Therefore, if the
$O_n$FT satisfies the extended analyticity, this must be true also for
$O_n$QG. If this is the case, the function $\Phi(\eta)$ must be
analytic in the whole $\eta$ plane with two cuts going from the points
\re{YLbps} to infinity, including the `shadow domain' 
which separates the HT and the LT wedges.

\section{Analysis of the matrix model explicit solution}
\label{section:sol}

In this section we show that the free energy extracted from the exact
solution of the discrete model does have the properties claimed in the
previous section.

\subsection{Lattice formulation on a planar graph}
\label{section:lattice}

The $O(n)$ loop model can be defined on any trivalent planar graph
$\Gamma$.  For any value of $n$, the partition function is equal to
the sum of all configurations of self- and mutually avoiding loops
with activity $n$ and open lines with activity $\vec H^2$,
\begin{equation}\label{looprONGG}
 \CZ_{_{O(n)}}(T, H; \Gamma)=\sum_{\text{polymers on }\Gamma}\;
 (1/T)^{\text{total length of the polymers}}\, n^{\#\text{loops}} \,
 \vec H^{2 \#\, \text{open lines}}.
\end{equation}
An example of such a configuration on a graph with the topology of the
disk is given in Fig.\ref{fig:Disk}.

In the $O(n)$ model on a dynamical lattice, the planar graph $\Gamma$
itself is treated as a statistical object.  The partition function of
the $O(n)$ model on a dynamical sphere of volume $\CN$ is defined as
the average of the partition function \re{looprONGG} over all graphs
$\G_N$ having $N$ vertices and the topology of a sphere,
   \be
   \la{micro-canonical-s}
\begin{aligned}
   \CZ_N^\sp (T, \vec H) & = \sum_{\Gamma_N} \CZ_{_{O(n)}}(T, \vec H;
   \Gamma_N)\, .
   \end{aligned}
   \la{defZNZN}
   \ee
Similarly, the partition function on a dynamical disk of volume $N$
and boundary length $L$ is defined as a sum over all planar graphs
$\Gamma_{N, L}$ with $N$ vertices and $L$ external lines, having the
topology of a disk,
   \be
   \la{micro-canonical-d}
\begin{aligned}
      \CZ_{N, L}^\di (T,\vec H) & = \sum_{\Gamma_{N, L}} \CZ_{_{O(n)}}(T,
      \vec H; \Gamma_{N,L})\, .
   \end{aligned}
   \la{defZNZNb}
   \ee
The disk partition function above is defined for the simplest boundary
condition with no special weight assigned to the boundary spins.  This
boundary condition means that the loops are repelled from the
boundary, as in Fig.  \ref{fig:Disk}.
   
We  are going   to compute the universal piece of the 
 specific free energy 
\be
\CF(T, \vec H)\defeq \lim_{N\to\infty} {\log   \CZ_N^\sp (T, \vec H)\over N},
\ee
 as well as  the   boundary
specific  free
energy  
\be \CF^B(T,\vec H) \defeq \lim_{N, L\to\infty}{ \CZ_{N, L}^\di (T,\vec H)- N
\CF(T,\vec H)\over L}.  \ee
For that it is easier first to evaluate the grand partition functions
\be \la{defSphpp}
\begin{aligned}
\CZ^\sp(\bmu, T, \vec H) &= \sum_{\CN=0}^\infty e^{- \bmu \CN}
\CZ^\sp_{\CN}(T, \vec H) \, ,\\
\CZ^\di(\bar\mu, \bmu_{_{\text{B}}} , T, \vec H) &= \sum_{\CN=0}^\infty
\sum _{L=0}^\infty e^{- \bmu \CN - \bmu_{_{\text{B}}} L} \
\CZ^\di_{\CN, L}(T, \vec H),
\end{aligned}
\ee
where $\bar\mu$ and $\bmu_{_{\text{B}}} $ are the lattice bulk and
boundary cosmological constants.  The large $N$ and $ L$ behavior of
the micro-canonical partition functions \re{defZNZN} and \re{defZNZNb}
is determined by the behavior of the grand partition functions near
the rightmost singularity in $\bar\mu$ and $\bmu_{_{\text{B}}} $,
which we denote respectively by $\bar\mu^c$ and $\bmu_{_{\text{B}}}
^c$.

The universal free energies are determined by the behavior of the
partition functions in the vicinity of the critical point $T=T^c,
H=0$, which is parametrized in terms of the renormalized couplings
\be \mu\sim \bar \mu - \bar\mu^c, \quad \mb \sim \bmb - \bmb^c\, \quad
t\sim T-T^c, \quad h\sim  |\vec H|.  \ee
All the information about the universal behavior is contained in the
singular parts of the partition functions, which we denote by $\CZ$
and $\CU$,
  \begin{align}
  \begin{aligned}
  \CZ^\sp(\bmu, T,\vec H) &= \CZ^\sp_{\text{reg}}(\bmu , T,\vec H) +
  \, \CZ(\mu, t, h), \\
    \CZ^\di(\bmu, \bmb, T,\vec H) &= \CZ^\di _{\text{reg}}(\bmu ,
    \bmb, T, \vec H) + \, \CU(\mu, \mb, t, h)\, .
    \end{aligned}
     \label{defcanZ}
    \end{align}
The regular terms describe degenerate configurations that diverge in
the large $N$ limit and should be thrown away in the renormalization
process.  The renormalized coupling constants $t, h, \mu, \mb$ can be
identified with coupling constants of Sect. \ref{section:dynamical}
up to numerical normalization factors.

\subsection{The solution}

We give the derivation of the analytic expressions for $\CZ$ and $\CU$
in appendix \ref{sec:MatrixH}.  The solution for the disk grand
partition function $\mathcal{U}$ is analytic in the boundary coupling
$\mb$ everywhere in the complex plane, except for a branch cut along
the interval $[-\infty, -M]$.  The function $M=M(\mu, t, h) $ gives
the position of the rightmost singularity of $\CU$ in the $\mb$
complex plane.  The renormalized boundary cosmological constant $\mb$
is defined so that at the critical point $\mu=t=0$ the singularity is
placed at the origin: $M(0,0,0)=0$.

The solution can be written conveniently in a hyperbolic
parametrization which resolves the branch point at $\mb=-M$.  The
boundary coupling $\mb$ is replaced by a hyperbolic parameter $\t$,
defined as
\be \mb\ \ = M\cosh(\t).  \ee
All the observables must be even functions of $\t$, and their
expression can be written solely in terms of hyperbolic cosines.  To
shorten the formulas, we introduce the notation
  \be C_a (\t) \defeq {M^{a } \over a} \cosh (a \t), \qquad a\in \IR.
  \ee
Then, up to a normalization of the coupling constants and the
partition functions,\footnote{ The signs cannot be changed by a
renormalization.  They follow from the requirement that the divergent
singular parts $(-\p_\mu)^3 \CZ$, $(-\p_x)^3 \CU $ and $(-\p_\mu)^2
\CU$ are positive.  (The singular parts that do not diverge contain a
finite regular positive part and do not need to be themselves
positive.)  } the function $M(\mu, t,h)$ is determined by the
transcendental equation
\be \la{srteqHa} \mu =      \frac{M^2}{2}-t \ \frac{M^{2(1- {1\over p}) }
}{2 (1-{1\over p} )}+h^2 {M^{-(1+{1\over p}) } \over 1+{1\over p} },
\ee 
and the derivatives of the disk partition function in $\mb$ and in
$\mu$ are written in the parametric form as
 \be \la{solutionn}
 \begin{aligned}
  -{\p \CU \over \p\mb}&= C_{{p+1\over p} }(\t) + t \, C_{{p-1\over p}
  }(\t) + {h^2} \ { M^{-{p-1\over p} }C_{{p-1\over p} }(\t)-
  M^{-{p+1\over p} } C_{{p+1\over p} }(\t)\over 2 \ [C_1(\t)]^2} \\
  {\p \CU\over \p \mu} &= { C_{1/p}(\t)} ,\\
   \mb &= C_1(\t).
\end{aligned}
\ee
Finally, the second derivative of the partition function on the sphere
is given by
 \be \la{defuofM} u \equiv -\p_\mu^2 \CZ = M^{{2/ p} }.  \ee One can
 eliminate $M$ from \re{srteqHa} and \re{defuofM} and write the
 equation of state for the susceptibility  $u =-\p_\mu^2 \CZ $,
\be \la{eqstateon} \mu=  {u^p\over 2} - t\ {u^{ p-1}\over 2 (1-{1\over
p} )} +h^2 \ {u^{-( p+1)/2 } \over 1+{1\over p} }\ .  \ee 

\subsection{ Microcanonical partition functions and analytic
expressions for the bulk and boundary specific free energies}

The specific free energy can be extracted from the leading exponential
behavior of the micro-canonical partition function when the volume
tends to infinity.  Knowing the grand canonical partition functions in
the thermodynamical limit, we can reconstruct the micro-canonical
ones.  In this limit, the discrete sums in \re{defSphpp} are replaced
by integrals with respect to the area $A$ and the length $\ell$:
 \begin{align}
\begin{aligned}
  \CZ(\mu, t, h) &= \int_0^\infty dA
\, e^{-\mu A} \  \CZ_A(t, h)
\\
 \CU(\mu, \mb, t, h) &=
\int_0^\infty dA \int _0^\infty d\ell \ e^{-\mu A- \mb \ell} \
\CZ_{A,\ell}(t, h)
 .  \la{Fixapp}
\end{aligned}
\end{align}
The partition functions for a given area and length are computed by
taking the inverse Laplace transforms.  With this definition the
regular part of the specific free energy is automatically subtracted.

For the partition function on the sphere, $\CZ_A$,  we find,
after integrating by parts,
\be\la{integrZA} \CZ_A(t, h) &=& {1\over A^2} \int {d \mu\over 2\pi
i} \ e^{\mu A} \ \p_\mu^2 \CZ(t, \mu, h) \no \\
&=&{1\over A^3}  \int  {du\over 2\pi i}   
\, 
\     e^{\mu (u) A} \  
,
\ee
where $\mu(u)$ is given by \re{eqstateon}.  The integral over $\mu$
goes along a contour going upwards and having all the singularities of
$\p_\mu^2 \CZ$ on the left.  The large $A$ asymptotics, given by the
contribution of the saddle point, is
\be\la{asymptAL} \CZ_A(t, h) \sim & A^{-3-1/2}\ e^{- \CF(t,h)\, A},
\ee
where $\CF(t,h)$ is by definition the specific free energy, evaluated
by the saddle point condition
 \be \la{defspecE} \CF(t, h) = -\mu(y), \quad (\p \mu/\p u)_{u=y} = 0
 .  \ee
 The exponent for the power  in the saddle-point result
 \re{asymptAL} corresponds to a pure gravity theory ($c_\text{matter}=0$).  
 This is because the asymptotic is evaluated for values of the area and 
 the length much larger than the correlation volume of the $O(n)$ spins,
 $A_\cor = 1/\CF(t, h)$.

In a similar way, the micro-canonical partition function on the disk
$\CZ_{A,\ell}$ is given by a double integral in $\mu$ and $\mb$.
   The  large $\ell$ asymptotics is determined by the rightmost singularity 
   of $\CU$ in the $\mb$ plane, which is the branch point at $\mb = - M$.
   This gives for the boundary free energy 
  \be\la{defspecEb} \Fb (t, h) = y^{p/2}, \quad (\p \mu/\p u)_{u=y} = 0.
  \ee
The bulk and the boundary specific free energies are intensive
characteristics of the system and as such do not depend on the global
properties of the random surface such as the number of handles or
boundaries.

From the equation of state  \re{eqstateon} one obtains the analytic
expressions for the specific free energies,
   \be \la{equ519}
   \begin{aligned} \CF(t, h) =& - { y^p\over 2} + t \, {y^{ p-1} \over
  2(1-{1\over p}) } - h^2\, {y^{-(p+1) /2}\over 1+{1\over p} }\\
    =& - \frac12
    y^{p-1}\left(\frac{3p+1}{p+1}y-\frac{p(3p-1)}{p^2-1}t\right) , \\
\Fb (t, h) =&  \ \ \ \ y^{p/2},
\end{aligned}
 \ee
where the saddle point value $u=y$ is the solution of
    \be
    \la{equ_saddle}
    y^{(3p-1) /2 }( y- \, t ) = h^2 .  \ee

 \subsection{The scaling functions $\CG_\high (\xi)$ and
 $\CG_\low(\z)$}

Let us first analyze the scaling function for the HT regime ($t>0$),
defined by \re{scalingHTd},
\be \CF(t, h) = t^{ p } \, \CGh (\xi )
, 
\quad \xi = h \, t^{-( {3p+{1})/4 }} .\ee
By \re{equ519} and \re{equ_saddle}   the scaling function
  is given in a parametric form by
\begin{align}
\begin{aligned}
\CGh &= \frac12 y^{p-1}\left(\frac{p(3p-1)}{p^2-1}-
\frac{3p+1}{p+1}y\right) 
 \\
    \xi ^2 \quad &= y^{(3p-1)/2 } \ \( y-1\) \, .  
    \la{eqxiofy}
\end{aligned}
\end{align}
 The solution can be expanded (see Appendix \ref{appendix:expansions})
 in a Taylor series in $\xi ^2$,
 \begin{equation}
\la{HTexp}
\CG_\high(\xi)=\frac{p}{2}\sum_{n=0}^\infty
{\dfrac{\G\left[\frac{3p+1}{2}n-p\right]}{
\G\left[\frac{3p-1}{2}n+2-p\right]} {(- \xi^2)^n \over
n!}}\simeq\dfrac1{2(p-1)}-\dfrac{p}{p+1}\xi^2+\cdots
\end{equation}
The expansion is convergent in the circle $|\xi |<\xi_c $, where 
\be
\xi
_{c}=
\, \textstyle{ \sqrt{2} \ {(3p-1)^{3p-1\over
4}\over (3p+1)^{3p+1\over 4}}} 
\ee
is determined by the position of the nearest singularity $ d\xi /d
y=0$.  The function $y (\xi)$ has in general an infinite number of
branch points, all at distance $\xi_c$ from the origin,
\be
\xi _{k,\pm} = \pm i \, e^{i\pi k {3p+1\over 2}} \xi_c, \quad k\in\IZ
. \la{branchxi} \ee
The two branch points on the physical sheet, $\xi_\pm^\YL = \xi_{0,
\pm}= \pm i \xi_c$, are of course the positions of the two YL
singularities.  Near the Yang-Lee branch points the scaling function
has the expected behavior of the YL model coupled to gravity, eq.
\re{YLpred}.

Looking at the integral representation \re{integrZA}, it is easy to
see that the YL edges appear as the result of condensation of zeroes
of the microcanonical partition function $\CZ_A(t,h)$.  When
$\xi$ is close to $\xi_+^\YL$ or to $\xi_-^\YL$, the second derivative
$\mu''(u)$ vanishes at the saddle point together with $\mu'(u)$ and
the integral is approximated by Airy integral.  When $|\xi|< \xi_c$,
the fixed area partition function has no zeroes and the power $3/2$
singularity of $\CGh(\xi)$ reflects the asymptotics of the Airy
function for large positive values of the argument.  On the other
hand, when $\xi$ is purely imaginary and $|\xi|>\xi_c$, the product of
the Airy functions associated with $\xi_+^\YL$ and $\xi_-^\YL$ has
zeroes with density
 \be
  \rho \sim A^{1/3} \sqrt{\xi^2 + \xi_c^2 },
  \ee
 which condense into the two YL cuts.
 
 We saw that the scaling function $\CGh (\xi )$ has the analytic
 properties anticipated in Sect.  \ref{subsec:GHT}, for which there
 was little doubt.  Now let us turn to the scaling function $ \CGl (\z
 )$, whose analytical properties were the subject of the speculations
 of Sect.  \ref{subsec:GTL}.

The scaling function $\CGl (\z )$ for the LT regime ($t<0$), defined by
\re{scalingHTd},
\be 
\CF(t, h) = (-t)^{ p } \, \CGl (\z )
,\quad \zeta = h\ (-t)^{-( {3p+{1})/4 }} ,\ee
is written in  parametric form as
\begin{align}
\begin{aligned}
       \CGl \  
&= - \frac12 y^{p-1}\left(\frac{p(3p-1)}{p^2-1}+
\frac{3p+1}{p+1}y\right) \, ,
 \\
    \z ^2 \quad  &= \ y^{(3p-1)/2  } \ \( y+1\)  
    \, .
\end{aligned}
 \la{eqxiofyLT}
\end{align}
 The asymptotics for small and large positive $\z$ have the form
 anticipated in Sect.  \ref{subsec:GTL},
      \be\la{asymptotxi} 
     \CGl(\z) =
 \begin{cases}
 \ \z^{{4 p\over 3p+1}} & \text{ if } \ \z \to \infty , \\
       \z^{{4(p-1)\over 3p- 1}} & \text{if }\ \z\to 0\, .\\
\end{cases}
 \ee
On the principal sheet, the cut relating the branch points at $\z=0$
and $\z=\infty$ is placed on the negative real axis.  The small $\z$
expansion of the bulk free energy is computed in Appendix
\ref{appendix:expansions}.  It is indeed of the form \re{ZZexp},
  \begin{equation}
  \begin{aligned}
 \CG_\low(\z)
 &=- {p(3p-1)\over 2(p^2-1)} \z^{\frac{4(p-1)}{3p-1} } +{1\over 2}
 \z^{\frac{4(p-1)}{3p-1}+\frac{4}{3p-1}} +{p\over 2(3p-1)}
 \z^{\frac{4(p-1)}{3p-1}+\frac{8}{3p-1}} + \dots \ .
 \end{aligned}
 \la{expGLz}
\end{equation}
 The coefficient $\tilde \CG_n$  of the series are given by
  \be
  \tilde \CG_0=0\, ;\qquad
  \tilde \CG_n={1\over n!}\ 
  \dfrac{p }{3p-1} \
 {\dfrac{\G\left[\frac{3p+1}{3p-1}n-\frac{p+1}{3p-1}\right]}{
 \G\left[\frac{2 }{3p-1} n+\frac{5p-3}{3p-1}\right]}}
 \qquad (n\ge 1).
 \ee
 The vanishing constant term reflects our convention for the
 integration constant in the derivation of the transcendental equation
 for $\mu$ in Appendix \ref{sec:Scalimit}.  A non-zero constant term
 is generated by a redefinition of the cosmological constant $\mu\to
 \mu + c |t|^p$.

\subsection{The  scaling function $\Phi(\eta)$}

Here we analyze the scaling function $\Phi(\eta)$ defined by
\re{defPhif},
\be \CF(t, h) = h^{ {4p\over 3p+1}} 
\, \Phi(\eta), 
\quad \eta= t\, h^{- { 4\over 3p+1}} .  \ee
Using \re{equ519} and \re{equ_saddle} we write the parametric
representation of $\Phi(\eta)$,
\be \la{Phieta}
\begin{aligned}
  \Phi&= \hf \textstyle{ {p(3p-1)\over p^2-1} }\ v^{-{p+1\over
  2}} -\hf \textstyle{ {2p-1\over p-1} }\ v^{p}
     \\
     \eta &= v - v^{-{3p-1\over 2}} \, .
\end{aligned}
\ee
The   function $\Phi(\eta)$ is analytic inside
 the disk
\be |\eta| < \eta_c, \qquad \eta_c = \xi_c^{-1/\kappa}=
\textstyle{{3p+1\over 2} \({3p-1\over 2}\)^{-{3p-1\over 3p+1}}},
\la{circleanalPhi} \ee
where it is given by the power series  
\re{Tayloretad}. For the coefficients $\Phi_n$ we find
(Appendix \ref{appendix:expansions})
 \begin{equation}
\begin{aligned}
\Phi _0 &=-\dfrac{3p+1}{2(p+1)} 
\, ;
\qquad
\Phi_n  = \dfrac{p}{3p+1}  
{(-1)^n\over n!} {\dfrac{\G\left[\frac{2n}{3p+1}
-\frac{2p}{3p+1}\right]}{ \G\left[\frac{4p+2}{3p+1}
-\frac{3p-1}{3p+1}n\right]}}  \qquad (n\ge 1).
\end{aligned}
\end{equation}
For general $p$ the function $\Phi(\eta)$ has an infinite number of
simple branch points, the solutions of the equation $dv/d\eta=0$,
\be \eta_j=e^{ 2 \pi i { 2j-1 \over 3p+1}} \ \eta_c, \qquad j\in \IZ.
\ee
There are no other singularities at finite $\eta$.  The cuts starting
at $\eta_j$ go along the radial direction and end at the branch point
at $\eta=\infty$, which is in general of infinite order.  The
principal sheet has two branch points, $\eta^\YL_+=\eta_0$ and
$\eta^\YL_-=\eta_1 $, which are the images of the YL branch points in
the $\xi$ plane, as shown in Fig.  \ref{fig:etaplaned}.  The two
branch cuts separate the sectors HTW
and LTW$\cup$SHD defined in Sect.  \ref{sect:Phieta}, in which the
scaling function has different asymptotics,
      \be\la{asymptoteta}
   \Phi(\eta) &\sim &
 \begin{cases}
\ -\ \eta ^p & \text{ if } \ \ \eta\to\infty, \ \eta \in \text{HTW} ,
\\
  - (-\eta)^{p+1\over 3p-1} & \ \text{if }\ \eta\to\infty, \ \eta \in
  \text{LTW}\cup \text{SHD}\, .\\
\end{cases}
 \ee

According to \re{Phieta}, the Riemann surface of $\Phi(\eta)$ is
symmetric under a rotation at angle ${4\pi\over 3p+1}$.
Therefore all sheets are rotated images of the principal sheet.  Let
us label the sheets by the integers $\dots, S_{-1}, S_0, S_1, \dots$,
where $S_0$ is the principal sheet.  Then the sheet $S_k$ has two cuts
going in the radial direction and starting at the points $\eta_k$ and
$\eta_{k+1}$.  For irrational $p$, the Riemann surface has an infinite
number of auxiliary sheets.

\begin{figure}
\begin{center}
   \includegraphics[width=11cm]{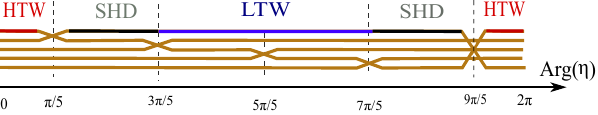}
\caption{\small A circular slice of the Riemann surface of $\Phi(\eta)
$ for $m=2$ $(p=3)$.  }
\label{fig:Phieta}
\end{center}
\end{figure}

Let us now consider the case when $p= 2m-1$ is odd integer, when the
matter QFT is a perturbation of a unitary CFT. The Riemann surface of
$\Phi(\eta)$ has $ {3p+1\over 2} =3m-1$ sheets and $3m-1$ simple
branch points $\eta_0, \dots, \eta_{3m-2}$.  At infinity, the $3m-1$
cuts join in a ramification point of order $ 3m-2$ where $\Phi\sim
\eta ^{ m/( 3m-2)} $.  If we cut all sheets along a circle with radius
larger $\eta_c$, the exterior part of the Riemann surface will split
into $2$ disconnected pieces as shown in Fig.  \ref{fig:Phieta}.
When analytically continuing from the HTW, one returns to the
principal sheet after circling once the origin.  When analytically
continuing from the LTW, one should circle $3m-2$ times around the
origin in order to return to the principal sheet.

 \subsection{The complex curve of the specific free energy}

The two scaling functions were obtained by eliminating $y$ from the
two equations \re{equ519} and \re{equ_saddle}, assuming that the
temperature coupling is positive or negative.  The equations allow to
consider the coupling constants $t$ and $h$ as non-restricted complex
variables and define the scaling functions in the HT and the LT
regimes as the restrictions to two different sheets of the same
analytic function $\CG(\xi)$, defined by \re{eqxiofy},
\begin{align}
\begin{aligned}
\CG \, &=\hf \textstyle{p (3p-1)\over p^2-1}\, y^{p-1} - \hf
\textstyle{3p+1\over p+1}\, y^p \\
    \xi ^2 &= y^{(3p-1)/2 } \ \( y-1\) \, . 
     \la{complexcurve}
\end{aligned}
\end{align}
In this sense one can speak of the complex curve of the specific free
energy.  When $p$ and $2\kappa = {3p+1\over 2}$ are integers, this is
an algebraic curve of genus zero.

The scaling function $\CG_\high(\xi)$ is the restriction of $\CG(\xi)$
on the principal (HT) sheet, while the scaling function $\CG^\low$ is
obtained as the analytic continuation of $\CG_\high(\z)$ below the YL
cuts,
 \be \la{Glowhigh} 
 \begin{aligned}
 \CG_\high(\xi) &= \CG(\xi), \qquad\qquad\qquad\qquad\qquad \xi\in
 \text{ HT sheet}, \\
 \qquad \CGl(\z)&= e^{\mp i\pi p} \, \CGh(e^{\pm i\pi{3p+1\over
 4}}\z), \ \qquad \z=e^{ \mp i\pi\textstyle{3p+1\over 4}
 } \xi\in \text{LT
 sheet}.
 \end{aligned}
  \ee

\begin{figure}[h]
         \centering
            \centering
           \includegraphics[width=10.0cm]{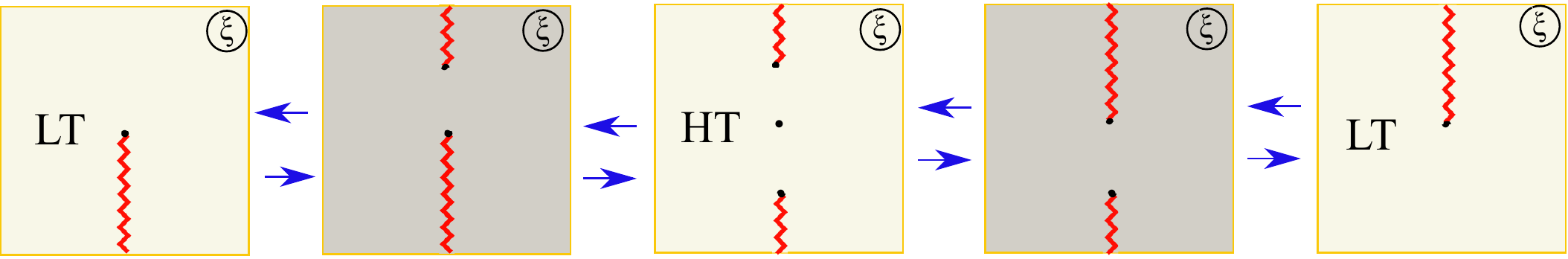} 
 \caption{ \small  The five sheets of the Riemann surface for the
   Ising model, $p=3$.  }
    \label{fig:SHEETS-Ising}
      \end{figure}

 Let us first remind the structure of the Riemann surface 
 of $\CG(\z)$ in the simplest case of the  Ising model, which has
 been analyzed in the unpublished work by Al. Zamolodchikov 
 \cite{AlZ-unp}.  When $p=3$,
 the Riemann surface represents a five-sheet cover of the complex
 $\xi$-plane.  The five sheets of the Riemann surface of the curve
 $\xi^2 = y^5 - y^4$ are shown in Fig.\ref{fig:SHEETS-Ising}.  There
 are two copies of the LT sheet, each of then connected to the HT
 sheet by an auxiliary sheet having two cuts.  After changing the
 variable $\xi\to \z = e^{\mp i 5\pi /4} \xi$, the Langer cut in the
 two copies of the LT sheet places itself on the negative axis in the
 $\z$-plane.
 
        \bigskip

For general $p$ the analytic continuation from the HT sheet can be
performed in four different ways, by going in clockwise/anticlockwise
direction beneath the upper/lower YL cut.  That is, the HT sheet is
connected in general to four different copies of the LT sheet, which
form an orbit of the $Z_2\times Z_2$ symmetry \re{symmetryRS} of the
meromorphic function $\CG(\xi)$.  This symmetry preserves the HT sheet
(or each of the copies of the HT sheet, if there are several of them),
but not the LT sheets.

 \begin{figure}[h]
         \centering
         \begin{minipage}[t]{0.4\linewidth}
            \centering
            \includegraphics[width=7.0cm]{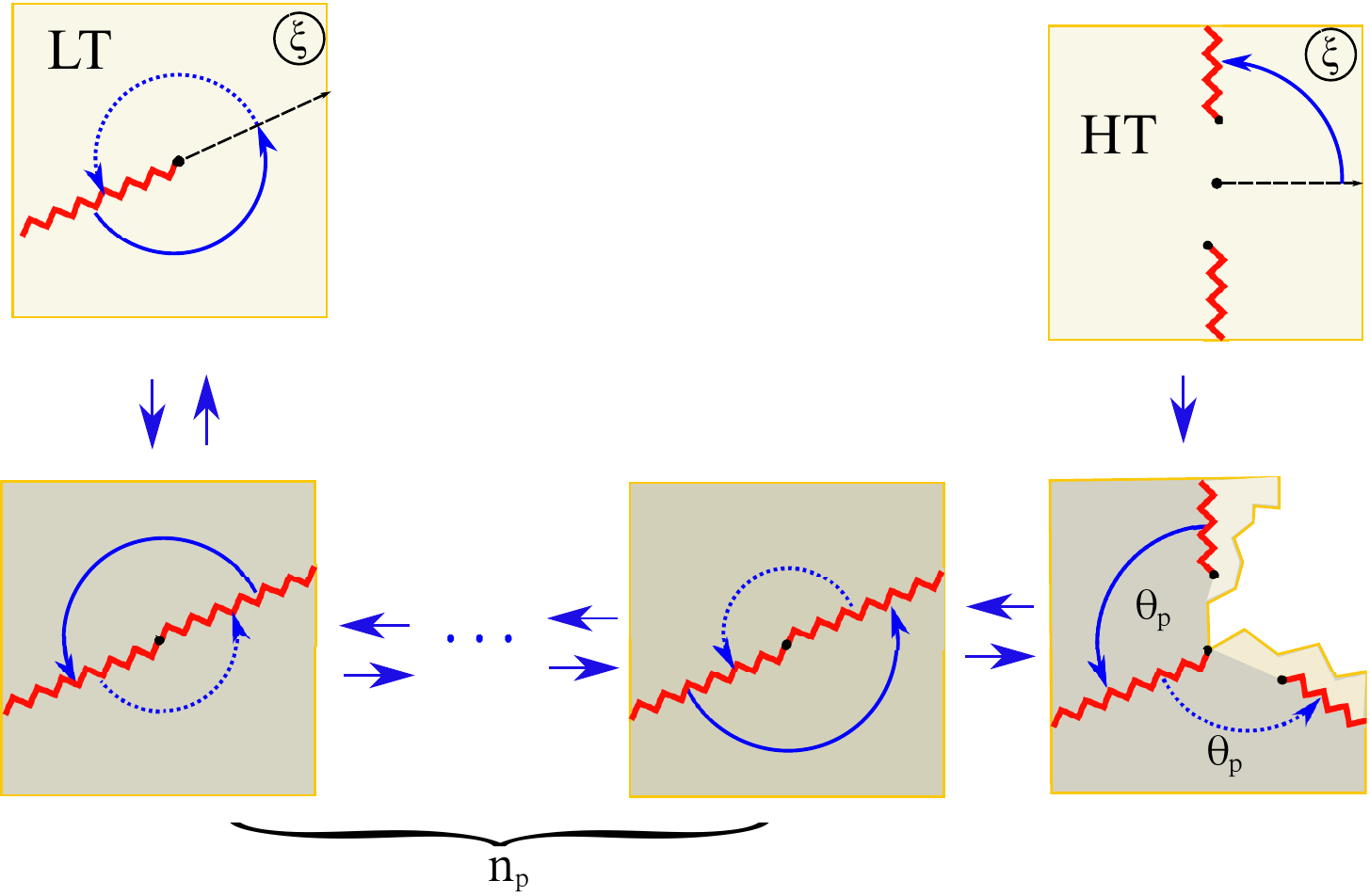} 
 \caption{ \small  Analytic continuation under the YL cut from the HT
   sheet to the LT sheet.} \label{fig:SHEETSxi}
         \end{minipage}%
         \hspace{2cm}%
         \begin{minipage}[t]{0.4\linewidth}
            \centering
            \includegraphics[width=3.5cm]{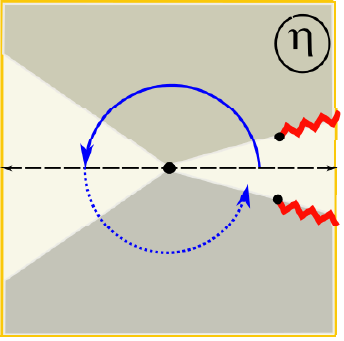}
 \caption{\small Analytic continuation under the YL cut from the HT
 sheet to the LT sheet.  }
\label{fig:SHEETSeta}
         \end{minipage}
 \vskip 0.2cm
      \end{figure}

In Fig.\ref{fig:SHEETSxi} we show the part of the Riemann surface
containing a path, starting from the real axis on the HT sheet, going
under the upper YL cut and continuing anti-clockwise, until it reaches
the positive axis ($\z \equiv e^{i \pi {3p+1\over 4}}\xi >0$) of the
LT sheet.  For convenience we place the Langer cut on the negative
$\z$-axis of the LT sheet.  The total angle swept by the path is
$\kappa \pi = {3p+1\over 4} \pi $.  We taylor the sheets of the
Riemann surface visited by the path according to the following
decomposition of the angular distance along the path,
\be
\la{npthp}
 \textstyle{3 p +1\over 4} \pi  
 = \hf \pi + \th_p+ n_p \pi +\pi
    \, ,
 \qquad (n_p \ \text{ integer}, 
 \ \  0< \th_p\le \pi).
 \ee
 This means that the second sheet has, besides the YL cut, a cut along
 the ray $ \arg(\xi) = \hf \pi + \th_p$, and possibly other cuts.  The
 remaining $n_p$ identical sheets have two cuts along the rays
 $\arg(\xi) = \hf \pi + \th_p$ and $\arg(\xi) = \hf \pi + \th_p+\pi$.
 The axis $\z>0$ of the LT sheet is rotated with respect to the
 positive axis of the $\xi$-plane at angle ${\pi\over 2}+\th_p$ if
 $n_p$ is odd, and ${\pi\over 2}+\th_p+\pi $ if $n_p$ is even.  The
 image of the path in the $\eta$-plane starts at the positive axis and
 ends at the negative axis, staying all the time in the upper half of
 the principal sheet of the function $\Phi(\eta)$, as shown in
 Fig.\ref{fig:SHEETSeta}.  If the path is continued further
 anti-clockwise (the punctured line), it will visit once again the
 $n_p$ sheets and then the second sheet, to reach (in general another
 copy of) the HT sheet.  The image of the path in the $\eta$-plane
 returns to the positive axis.

 If one continues in the same way, the same pattern will reappear,
 rotated at angle $2\th_p$, and so on.  The symmetry $\xi\to
 e^{2i\th_p} \xi$ of the Riemann surface of $\CG(\xi)$ reflects the
 symmetry $\eta\to e^{2i\pi}\eta$ of the Riemann surface of
 $\Phi(\eta)$.  For rational values of $p$, the path will eventually
 return to the starting point after a finite number of turns around
 the origin.

 Let us   consider  some particular cases.

\bigskip

$\bullet$ Polymers ($n=0, \ p=2$).

\begin{figure}[ht]
         \begin{minipage}[t]{0.4\linewidth}
            \centering
 \includegraphics[width=5.0cm]{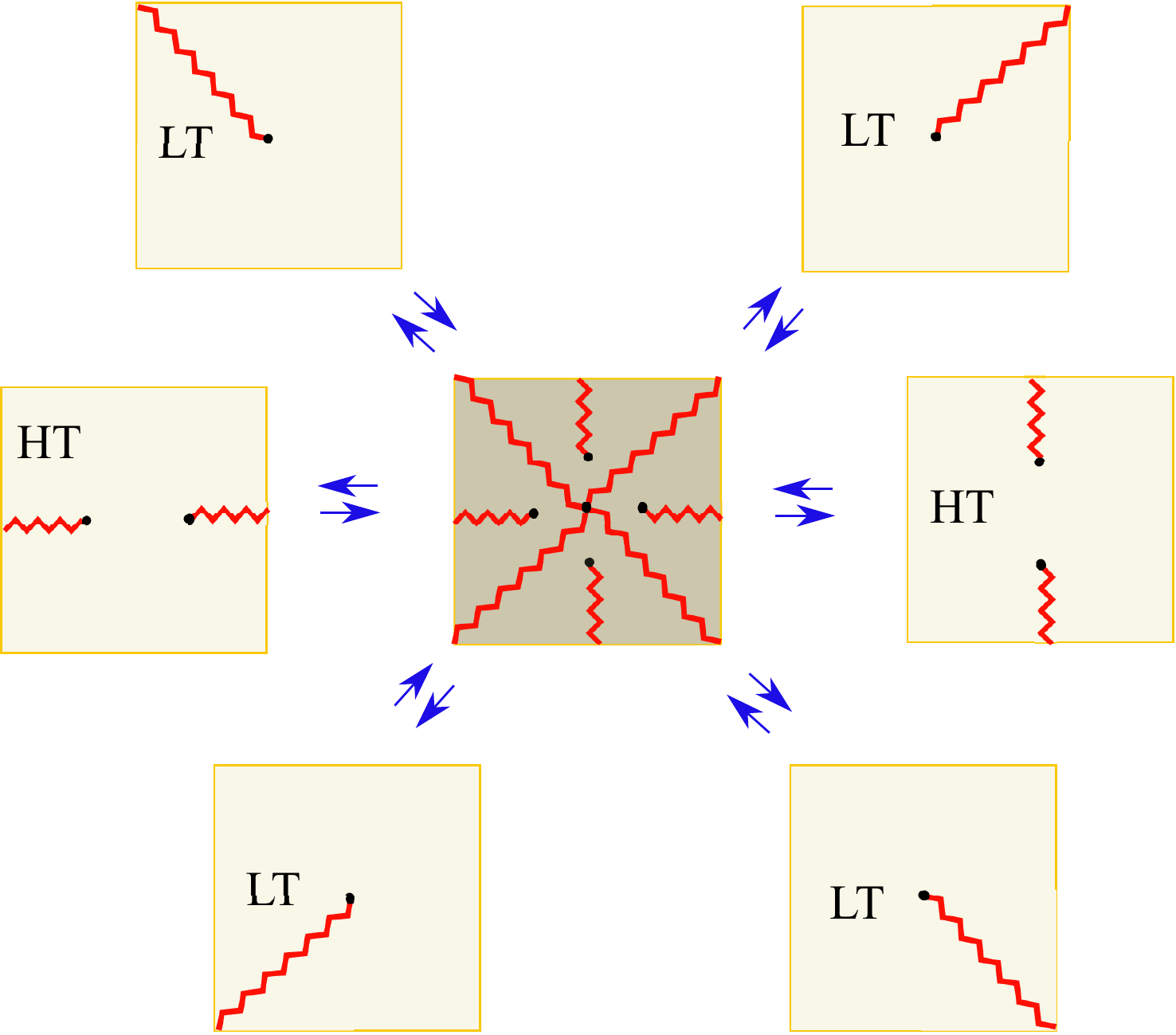} 
   \caption{\small The seven sheets of the Riemann surface for linear
   polymers, $p=2$ ($n_p=0$).  }
\label{figure:SHEETS-PG}
 \end{minipage}
   \centering  
  \end{figure}

The partition function is that of a grand canonical ensemble of open
linear polymers.  The Riemann surface for curve $\xi^2 =
y^{7/2}-y^{5/2}$ has 7 sheets, depicted in Fig.\ref{figure:SHEETS-PG}.
There are two copies of the HT sheet and four copies of the LT sheet.
There is only one connecting sheet on which the function $y(\xi)$ has
four simple branch points at $\pm \xi_c, \pm i \xi_c$, as well as at
$\xi=0$ and $\xi=\infty$.

 %
 \begin{figure}[h]
    \begin{minipage}[t]{0.4\linewidth}
            \centering
  \includegraphics[width=6.5cm]{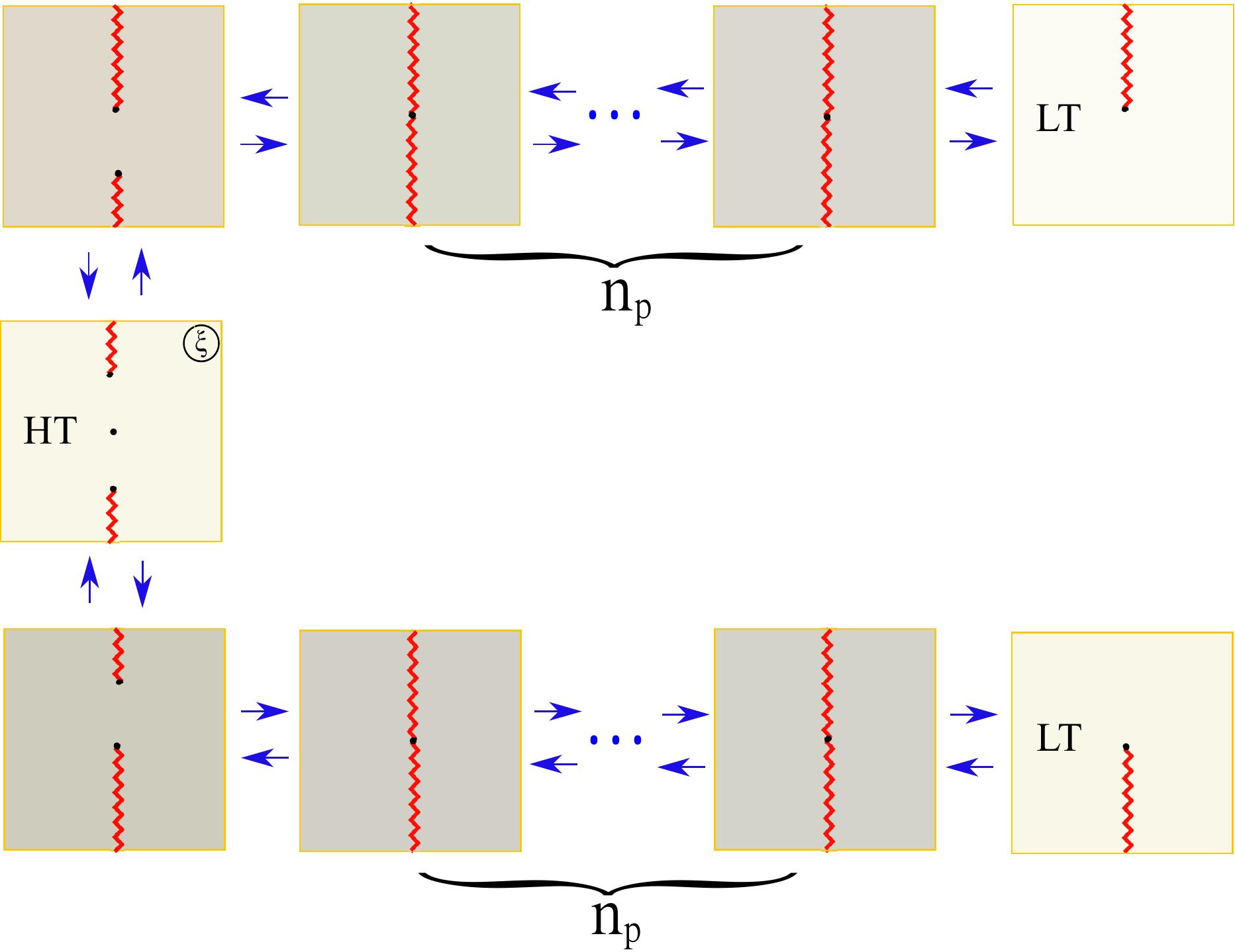} 
   \caption{\small The Riemann surface for $p=2m-1$ with $m$ even.
   Here $n_p= {3\over 2} m -3$, $\th_p=\pi$.}
\label{figure:SHEETS-unitary}
 \end{minipage}
    \hspace{1cm}
  \begin{minipage}[t]{0.4\linewidth}
            \centering
  \includegraphics[width=7.9cm]{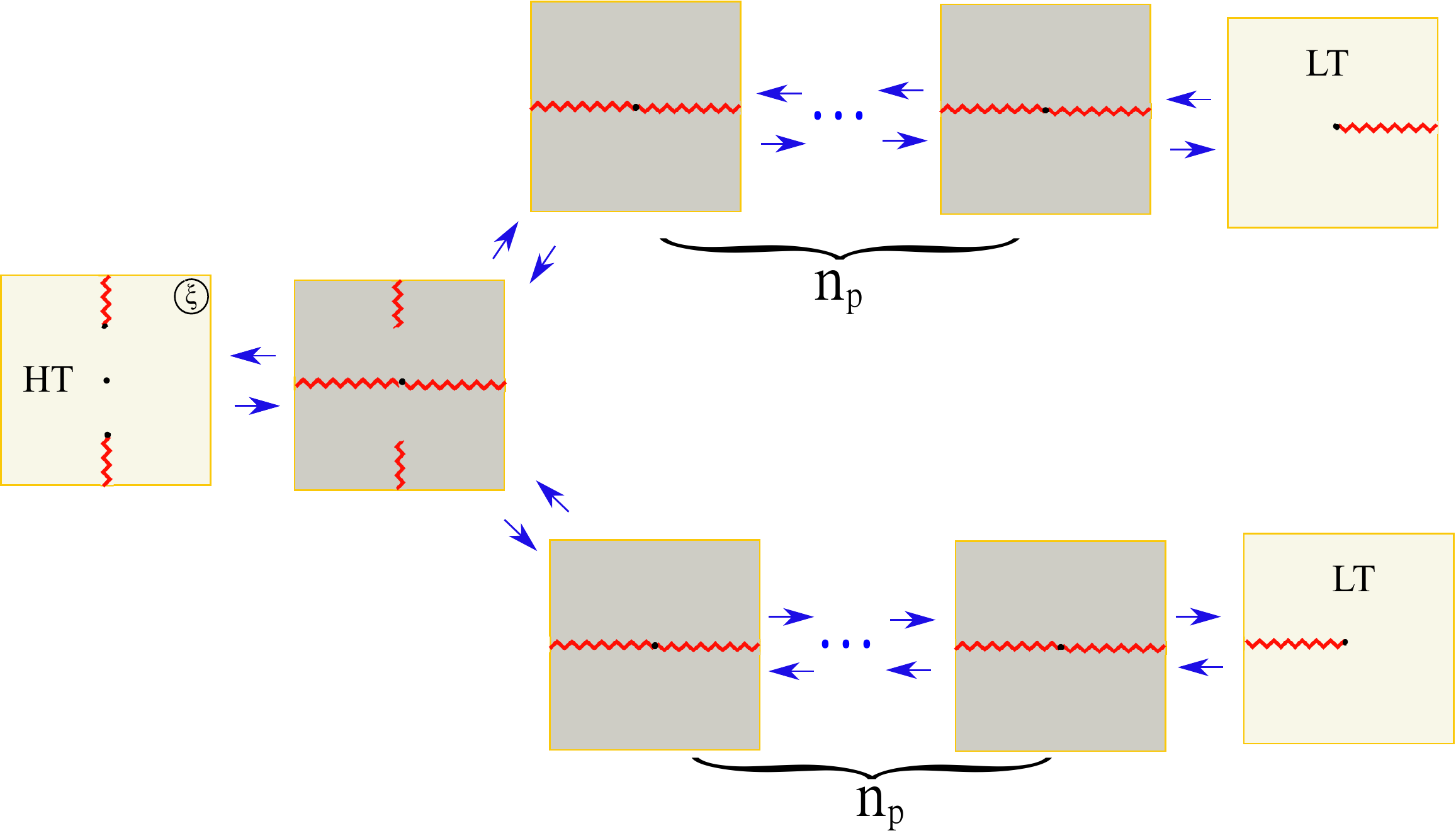} 
   \caption{\small Riemann surface for $p=2m-1$ with $m$ odd.  Here
   $n_p= {3m - 5 \over 2} $, $\th_p=\hf \pi$.}
\label{figure:SHEETS-nonunitary}
 \end{minipage}
  \end{figure}

$\bullet$    $p= 2m-1, \quad m\in\IN $. 
   
In this case the critical theory is a unitary CFT and the operator
coupled to the magnetic field is the primary field $\Phi_{m,m}$.  When
$p $ is odd integer, the equations \re{complexcurve} become algebraic.
The function $\CG(\xi)$ has, apart of $\xi=0$ and $\xi=\infty$, only
two branch points, $\xi = \pm i \xi_c$.  The Riemann surface has
$3m-1$ sheets, which include one HT sheet and two copies of the LT
sheet.  The $3m-1$ sheets of the Riemann surface are shown in
Fig.\ref{figure:SHEETS-unitary}, for $m$ even, and in
Fig.\ref{figure:SHEETS-nonunitary}, for $m$ odd.  In the first case
$n_p= {3\over 2} m -3$ and $\th_p=\pi$, while in the second case $n_p=
{3m - 5 \over 2} $ and $\th_p= \hf \pi$.

\section*{Acknowledgments}
I.K. thanks S. Lukyanov and H. Saleur for illuminating  discussions.
 J.E.B. acknowledges the warm hospitality of the CEA-Saclay, KEK, Yukawa Institute and Tokyo University, and  thanks A. Kuniba
and  K. Hosomichi for useful discussion.  This work is partially
supported by the National Research Foundation of Korea (KNRF) grant
funded by the Korea government (MEST) 2005-0049409 (JEB).

\appendix

\section{ Matrix model solution for $O(n)$ spin system in magnetic field}
\label{sec:MatrixH}

\subsection{Mapping to a matrix model and loop equation for the resolvent}

\begin{wrapfigure}{r}{0.6\textwidth}
  \vspace{-20pt}
  \begin{center}
    \includegraphics[width=0.55\textwidth]{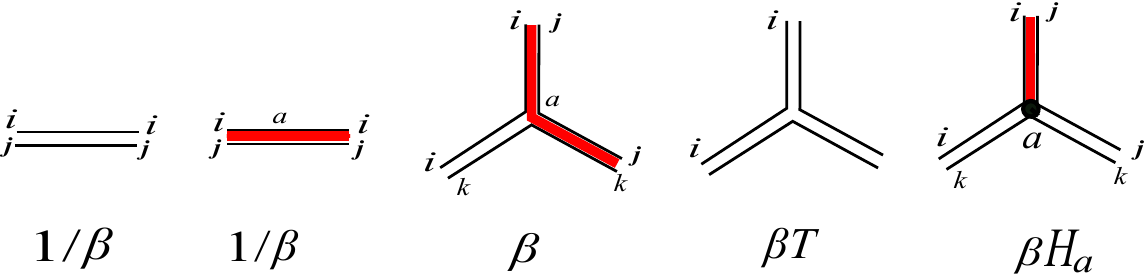}
  \end{center}
  \vspace{-20pt}
  \caption{\small Feynman rules for the $O(n)$ matrix model in the
presence of a magnetic field.  The magnetic term $\vec H\cdot \vec{\bf
Y}$ can be seen as a source for tadpoles.}
\label{fig:Feynman}
\end{wrapfigure}

The $O(n)$ matrix model \cite{Kostov:1988fy}, with an extra term
taking into account the magnetic field, is the zero-dimensional planar
field theory whose Feynman graphs are built according to the Feynman
rules shown in Fig.\re{fig:Feynman}.  The partition function of the
matrix model represents the integral over one hermitian matrix ${\bf
M}$ and a $O(n)$ vector $\vec {\bf Y}$ whose components are hermitian
matrices,
\begin{equation}\label{partfM}
{Z}_N \sim \int d{\bf M}\, d^n{\bf Y} e^{\beta\tr[ - {1\over 2} \Tr
{\bf M}^2 - {1\over 2} \Tr \vec {\bf Y}^2 + {T\over 3} {\bf M}^3 +
{1\over 2} {\bf M}\vec{\bf Y}^2 + \vec H\cdot \vec{\bf Y}]}\,,
\end{equation}
At this stage the number $n$ is considered integer.  Expanding the
free energy of the matrix model as a sum over Feynman diagrams, we
obtain a sum over all polymer configurations on the disk as the one
shown in figure \ref{fig:Disk}.  The weight of these diagrams is given
by
\begin{equation}
N^{2-2g}\, \left(\dfrac\b{N}\right)^{-\#\text{vertices}/2}\,
T^{-L_{\rm tot}}\, n^{\#\text{loops}} \, H^{2 \#\, \text{open lines}},
\end{equation}
where $N$ is the size of the matrices and $g$ is the genus of the
`fat' planar graph.  The functions defined in \re{defSphpp} are
expressed in terms of the matrix model as
\be 
\begin{aligned}
\la{ws-matrix}
\CZ^\text{sphere}(\bar{\mu},T,H)&=\lim_{N,\b\to\infty}\dfrac1{N^2}\log
Z_N,\\\
\CZ^\di(\bar\mu,   \bmb , T, H) 
&=\lim_{N,\b\to\infty}\dfrac1{N^2}
\<\Tr \log\(1- e^{- \bmb} {\bf M}\)\>,
\qquad
\beta/\CN= e^{2\bmu} ~~ \text{fixed}.
\end{aligned}
\ee

 It is useful to shift the matrix variable as $ {\bf M} = {\bf X}
 +{1\over 2} {\bf I}$.  Integrating over ${\vec{ \bf Y}}$, we write
 the partition function \re{partfM} as an integral over the
 eigenvalues of ${\bf X}$,
\be\label{partfXX} {Z}_N &\sim& \int \prod_{i=1}^N {d x_i \over ( 2
x_i)^{n/2}} \, e^{- \b V(x_i)} \prod _{i<j} {(x_i-x_j)^2\over
(x_i+x_j)^n}.  \ee with (after rescaling of $H$)
\begin{equation}
V(x)= -\frac T3\Big(x+\tfrac12\Big)^3 +\frac12\Big(x+\tfrac12\Big)^2 +
{H^2 \over x} .
\end{equation}
 The basic observable in the matrix theory, the resolvent
 \be\label{resW} W(x )= \frac1\beta\left\langle\text{tr} \frac1{x
 -{\bf X}} \right\rangle , \ee
is equal, by \re{ws-matrix}, to the derivative of the disk partition
function with respect to the boundary cosmological constant.  One can
check that the critical value of the later is $e^{\mb} = 1/2$, so that
the spectral parameter $x$ in \re{resW} is equal, up to a
normalization, to the renormalized boundary cosmological constant,
$x=\mb$.

The resolvent $W(x )$ splits into a singular part $w(x )$ having a
branch cut on the real axis, and an analytic part $W_\reg(x )$:
\be \la{WregWsin} W(x ) \defeq W_\reg(x ) + w(x ) \, .  \ee
The singular part of $W$ is proportional to the derivative of the disk
free energy in the continuum limit $\CU(\mb)$, defined \re{defcanZ},
\be w(x) = - \CU'(x).  \ee
The regular part is given by
 \be W_\reg(x )&=& \frac{2 V'(x )-n V'(-x )}{4-n^2} =-{H^2\over
 (2+n)\, x ^2}-a_0-a_1x -a_2x ^2
 .
 \label{defGx}
\end{eqnarray}
The function $w(x)$ satisfies a quadratic identity that can be
obtained using either the loop equation technique or the saddle point
approach,
\begin{equation}\label{funceqWh}
{ w^2(x )+w^2(-x )+nw(x )w(-x )\over 4- n^2} = c_{-4} \, x ^{-4} +
c_{-2} \, x ^{-2} + c_0 +c_2x ^2+c_4x ^4 \, .
\end{equation}
The coefficients $c_k$ are functions of $ T$, $\bar\mu, H$,
$W_1=\b^{-1} \langle\text{tr}X\rangle$, $W_{-1} = \b^{-1} \langle\tr
X^{-1}\rangle$ and $W_{-2} =\b^{-1} \langle\tr X^{-2}\rangle$ and can
be evaluated from the large-$x $ and small-$x $ asymptotics
  \be \label{asymG} w(x ) &=& a_2 x ^2+a_1x +a_0
  +\frac{e^{2\bar\mu}}{x } +\frac{ W_1}{ x ^2}+\frac{H^2}{(2+n)x
  ^2}+O(x ^{-3}), \qquad x \to\infty \no \\
 w(x ) &=& { b_0\over x ^2} - W_{-1} - W_{-2} \, x +a_1x +O(x ^{2}),
 \qquad x \to 0 \ee
in \re{funceqWh}.  Obviously
\be c_{-4}= {b_0^2 \over 2-n}, \quad b_0 = {H^2 \over 2+n}.  \ee
The solution of the loop equation \re{funceqWh} with the asymptotic
\re{asymG} is given by a meromorphic function with a single cut
$[a,b]$ on the first sheet, with $a<b<0$.  The coefficient $c_{-4} $
is proportional to $ H^4$, $c_2$ depends only on $T$ and $c_4$ is a
constant.  The coefficients, $c_{-2}$ and $c_0 $ depend on $ T$,
$\bar\mu, H$ directly and through the unknown quantities $W_1$,
$W_{-1}$ and $W_{-2}$.  Fortunately, we do not need to compute these
coefficients explicitly in order to find the solution in the scaling
limit, as we will see in the next section.

%
\subsection{Solution in the scaling limit}
\label{sec:Scalimit}

We will solve the loop equation in the vicinity of the double
singularity $T=T_\text{crit}, \ H=0$ and $x =0,\ \bar\mu=\bar\mu^c$.
This vicinity is parametrized by the renormalized couplings $t\sim
T-T_\text{crit},\ h\sim H$ and $\mu_B\sim x ,\
\mu\sim\bar\mu-\bar\mu_c$.  We will use the parametrization \re{parnp}
of the dimensionality $n$, with  $1/p=\th$:
 \be n= 2\cos(\pi\th), \quad \th= {1 / p}.  \ee
We will choose a specific normalization of the solution which simplify
the formulas.  We normalize $t$ and $h$ in such a way that the
constants $b_0$ and $c_2$ are given by
  \be b_0 = { \sqrt{ 2-n} \over 2(1-\th)} h^2 , \quad
  c_2=\frac{1+\th}{1-\th}t.  \ee
Here we used the fact that $c_2$ depends only on $T$.Once this
normalization is fixed, the quadratic identity \re{funceqWh}
determines the solution uniquely.

In the continuum limit, the left bound of the interval supporting the
eigenvalue density is sent to $-\infty$, and we can use an hyperbolic
parametrization to resolve the branch cut of the resolvent on
$[-\infty,-M]$,
\begin{equation}\label{defxtau}
 \mb\equiv x= M\cosh\tau\, .
\end{equation}
The solution will be expressed in terms of the functions
\be C_a (\t) = {M^{a}\over a} \cosh (a \t), \qquad a = k\pm \th, \ \
k=\text{odd integer},\ee
 which form a complete set in the space of functions having the
 required analytic properties.  In particular they satisfy the shift
 equation \be C_\nu(\t+i\pi)+C_\nu(\t-i\pi)+nC_\nu(\t)=0\ee also
 obeyed by the resolvent $w(\tau)$.  Note also that the
 parametrization \re{defxtau} can be written as $x= C_1(\t)$.  In
 terms of the parameter $\t$, the functional equation \re{funceqWh}
 becomes a shift equation for the entire function $w(\tau)\equiv
 w[\mu_B(\tau)]$:
\begin{equation}\label{feqWtauh}
 {w^2(\tau+i\pi) +w^2(\tau)+ n\,w(\tau+i\pi)w(\tau) \over 4\sin^2(\pi
 \th) } = {c_{-4}\over [C_1( \t)]^4} + {c_{-2}\over [C_1( \t)]^2} +
 c_0 +c_2 \, [C_1( \t)]^2.
\end{equation}
It is noted that the term proportional to $c_4$ is dropped in the
renormalization process.

From the scaling dimension of the resolvent, $w\sim \mu^{(1+\th)/2}$
(by definition $\mu$ has scaling dimension $1$) it follows that the
r.h.s scales as $\mu^{1+\th}$.  From here, one finds the scaling of
the coefficients:
\be h\sim \mu^{3+\th\over 4},\quad t\sim\mu^{\th},\quad c_0\sim
\mu^{1+\th}, \quad c_{-2}\sim \mu^{2+\th}.  \ee

We will try to satisfy the functional equation \re{feqWtauh} by the
following ansatz,
 \be \label{xofcthd} w(\tau)&=& C_{1+\th}(\t) + t \, C_{1-\th}(\t) +
 h^2 {b_+ C_{1+\th}(\t) + b_- C_{1-\th}(\t) \over [C_1(\t)]^2 } \ee
The coefficients $b_\pm$ are fixed by the two leading terms in the
small $x $ behavior \re{asymG},
which can be transposed into conditions on $w(\t)$ near the point $\t=
\pm i \pi /2$,
\be b_\pm =\mp \frac12M^{-(1\pm \th)} .  \ee
For this solution the coefficients $c_{-2}$ and $c_0$ are \be
\la{ccMM}
\begin{aligned}
&c_{-2} =- \frac{h^2}{4(1-\th^2)}\left[ h^2M^{-2}+{2\over 1+\th}
M^{1+\th}-{2\over 1-\th}\, t\, M^{1-\th}\right]\ ,\\
&c_0=\dfrac14 M^{2+2\th}+ {1\over 2(1-\th^2)}\left(h^2M^{-1+\th} (1-
t\, M^{-2\th})-tM^2\right)+\dfrac14 \dfrac{1 }{(1-\th)^2}
\,t^2M^{2-2\th}.
 \end{aligned}
\ee
where $M=M(\mu, t, h)$.
 
We cannot determine the function $M(\mu, t, h)$ by comparing \re{ccMM}
with the r.h.s. of \re{funceqWh}, because the information about the
behavior at large $x$ is lost after taking the continuum limit.
Instead, we will make use of the fact that the derivative
$\partial_\mu w(x)$ does not depend on the potential and have, up to a
normalization, a standard form \cite{Kostov:2006ry}
\be\la{dmuwx} \partial_\mu w(x) =- M^{-1+\th}\ {\sinh \th \t\over
\sinh\t} .  \ee
Comparing this expression with the derivative of the solution
\re{xofcthd} with respect to $\mu$ at fixed $x$,
  \be\la{dGc} \partial_\mu w|_x = - \partial_\mu M\(M^\th - t
  M^{-\th}- h^2\, M^{-3}\) \frac{\sinh\theta\tau}{\sinh\tau}, \ee
 yields a compatibility condition, which can be written, after
 integration, as a transcendental equation for $M$:
\be
\la{bentr} \mu 
= \frac{M^2}{2}-t \ \frac{M^{2-2 \theta }
}{2-2\th}+h^2 {M^{-1-\theta } \over 1+\th}  . 
 \ee
The cosmological constant is determined up to  an arbitrary 
 integration constant
$ \sim t^{1/\th}$.   
 
The loop mass $M$ is proportional to the second second derivative $u =
-\p_\mu^2 \CZ$ of the partition function on the sphere in the
continuum limit, defined in \re{defcanZ}.  There is a simple heuristic
argument to see that.  The derivative $-\p_\mu \CU(\mb)$ is the
partition function on the disk with one marked point in the bulk.  In
the limit $\mb\to\infty$ the boundary length vanishes and the boundary
can be replaced by a point.  The leading term in the limit
$\mb\to\infty$ is therefore proportional to the second derivative $
\p_\mu^2 \CZ$.  Expanding at $\mu_B\to\infty$ we find
 \be - \p_\mu \CU \sim - \mb^{\th} + M^{2\th } \, \mb^{-\th} + \text{
 lower powers of } \mb \ee
 (the numerical coefficients are omitted).  We conclude that the
 second derivative of the sphere partition function is
\be u \equiv -\p_\mu^2 \CZ = M^{2\th}.  \ee Now we can write the
equation of state for the $O(n)$ model on the sphere in its final form
\begin{eqnarray}\label{eqnIst}
\mu = {u^{1\over\th} \over 2} - t\, {u^{ {1-\th \over\th} }\over
2-2\th} + h^2 \ {u^{-{1+\th\over 2\th}} \over 1+\th}, \quad
u=-\p_\mu^2 \CZ.
\end{eqnarray}
 Of course, this equation implies certain normalization of $\mu$ and
 $\mb$.

\section{Series expansions of the scaling functions}
\la{appendix:expansions}

We have to deal with the transcendental equations of the form
\begin{equation}
\la{Transcen} \mu=z+\l\, z^{ \a},\quad \a>1.
\end{equation}
The solution of this equation as a power series can be derived using
the Lagrange inversion theorem,
\begin{equation}
\label{sol_al}
\begin{aligned}
z^{ \g}&=\g\sum_{n=0}^\infty{\dfrac{(-1)^n\l^n}{n!}\dfrac{\G[\alpha
n+\g]}{\G[(\alpha -1)n+\g+1]}\ \mu^{\g+n(\alpha -1)}} \\
&= \mu^\g -\g \ \l \mu^{\g +\a-1}+ {\g(2\a-1+\g)\over 2}\ \l^2 \mu^{\g
+ 2(\a-1)}+\dots .
\end{aligned}
\end{equation}

\paragraph{HT expansion:}
We would like to find the series expansion in $\xi$ of $\CG_\high$,
defined by the eq.  \re{eqxiofy}.  It is easier to first consider the
derivative
\begin{equation}\label{dCG}
\dfrac{d\CG_\high}{d\xi^2}=-\dfrac{p}{p+1}\ y^{-(p+1)/2}, \quad
\xi^2=y^{(3p+1)/2}-y^{(3p-1)/2}.
\end{equation}
 After changing the variable to $y=1+f$, we write these equations as
\begin{equation}
\xi^{\frac{2(p+1)}{3p-1}}\dfrac{d\CG_\high}{d\xi^2}=
-\dfrac{p}{p+1}f^\frac{p+1}{3p-1}, \quad
\xi^{\frac{4}{3p-1}}=f^\frac2{3p-1}+f^\frac{3p+1}{3p-1}.
\end{equation}
We can now use the result \re{sol_al} with $\mu=\xi^\frac4{3p-1}$,
$\l=1$, $z=f^{2\over 3p-1}, \alpha = {3p+1\over 2}\text{ and
}\g={p+1\over 2} $ to write
\begin{equation}
\dfrac{d\CG_\high}{d\xi^2}=-\dfrac{p}{2}
\sum_{n=0}^\infty{\dfrac{\G\left[\frac{3p+1}{2}n+\frac{p+1}{2}\right]}{
\G\left[\frac{3p-1}{2}n+\frac{3p+1}{2}\right]} {(- \xi^2)^n \over
n!}}.
\end{equation}
Finally, integrating once with $\CG_\high (0)=1/(2(p-1))$ we get
\begin{equation}
\CG_\high(\xi)=\dfrac{p}{2}\sum_{n=0}^\infty
{\dfrac{\G\left[\frac{3p+1}{2}n-p\right]}{
\G\left[\frac{3p-1}{2}n+2-p\right]} {(- \xi^2)^n \over n!}}.
\end{equation}

\paragraph{LT expansion:}
For the derivative of the scaling function $\CG_\low(\z)$, defined by
\re{eqxiofyLT}, we find
\be {d\CG_\low \over d\z^2}= -{p\over p+1} \, y^{-{p+1\over 2}},
\qquad \z ^2 = y^{(3p-1)/2 } \ \( y+1\) \, .  \ee
 The equation for $\z$ is equivalent to \re{Transcen}, with $\l=1$,
 $\mu=\z^2$, $z=y^{3p-1\over 2}, \alpha = {3p+1\over 3p-1}, \g=-
 {p+1\over 3p-1} $, and by \re{sol_al} we get, after integrating over
 $\zeta^2$,
  \begin{equation}
  \begin{aligned}
\CG_\low(\zeta)&= \dfrac{p }{3p-1}\sum_{n=0}^\infty
{\dfrac{\G\left[\frac{3p+1}{3p-1}n-\frac{p+1}{3p-1}\right]}{n!\
\G\left[\frac{2 }{3p-1} n+\frac{5p-3}{3p-1}\right]}
\z^{\frac{4(p-1)}{3p-1}+\frac{4n}{3p-1}}}.
 \end{aligned}
\end{equation}
The other expansions for the sheets having a branch point at $\zeta=0$
are obtained by considering the other roots of the function
$\zeta^{1/(3p-1)}$.

\paragraph{Expansion at infinity ($\eta$-plane):}

Again we compute the derivative
\be { d\Phi\over d\eta}= \frac{p \ y^{-1+p}}{2 (p-1)} \ee
Tn order to use eq.  \re{sol_al} we set $y^{(3p+1)/2}=1+f$.  In the
new variable
\begin{equation}
\eta^{-\frac{3p+1}{3p-1}}=f^{-\frac{3p+1}{3p-1}}+f^{-\frac2{3p-1}},
\end{equation}
 and for the derivative of $\Phi$ we find, using that
 $y=(1+f)^{2/(3p+1)}=\eta^{-2/(3p-1)}f^{2/(3p-1)}$,
\begin{equation}
\dfrac{d\Phi}{d\eta}= \dfrac{p}{2(p-1)}\eta^{-\frac{2(p-1)}{3p-1}}
f^\frac{2(p-1)}{3p-1}.
\end{equation}
Applying \re{sol_al} with $\mu=\eta^{-\frac{3p+1}{3p-1}}$, $\l=1$,
$z=f^{-\frac{3p+1}{3p-1}}, \alpha =\frac{2}{3p+1},\text{ and
}\g=-\frac{2(p-1)}{3p+1} $ we get
\begin{equation}
\dfrac{d\Phi}{d\eta}=-\dfrac{p}{3p+1}\sum_{n=0}^\infty
{\dfrac{\G\left[\frac{2n}{3p+1}
-\frac{2(p-1)}{3p+1}\right]}{n!\G\left[\frac{p+3}{3p+1},
-\frac{3p-1}{3p+1}n\right]}(-1)^n\eta^{n}}
\end{equation}
and finally integrating this expression with the constant
$\Phi(0)=-(3p+1)/(2(p+1))$,
\begin{equation}
\Phi(\eta)=-\dfrac{3p+1}{2(p+1)}+\dfrac{p}{3p+1}\sum_{n=1}^\infty
{\dfrac{\G\left[\frac{2n}{3p+1}-
\frac{2p}{3p+1}\right]}{n!\G\left[\frac{4p+2}{3p+1}
-\frac{3p-1}{3p+1}n\right]}(-1)^n\eta^{n}}
\end{equation}
 The branches of $\Phi(\eta)$ for finite $\eta$ are in relation with
 the branch points of $y(\eta)$, solutions of $y^{3p+1\over 2} = -
 \hf(3p-1)$:
\be y_k = e^{ 2\pi i { 2k-1\over 3p +1}} \ |\hf(3p-1)|^{2\over 3p+1},
\quad k= 0, \pm 1, \dots , \ee
which gives for the positions of the branch points in the $\eta$ plane
\be \eta_k=e^{ i \pi (2k-1)/2\kappa} \ \eta_c, \qquad \eta_c =
\left(\frac{3 p-1}{2}\right)^{\frac{2}{3 p+1}} \frac{3 p+1}{3p-1}.
\ee
%

 

 \providecommand{\href}[2]{#2}\begingroup\raggedright\endgroup

\end{document}